\documentclass[fleqn,usenatbib]{mnras}

\usepackage[T1]{fontenc}

\DeclareRobustCommand{\VAN}[3]{#2}
\let\VANthebibliography\thebibliography
\def\thebibliography{\DeclareRobustCommand{\VAN}[3]{##3}\VANthebibliography}

\usepackage{graphicx}	
\usepackage{amsmath}	
\usepackage{amssymb}	

\usepackage{newtxtext,newtxmath}

\title[Nucleosynthesis of the most metal-rich AGB stars]{The most metal-rich stars in the universe: chemical contributions of low and intermediate mass asymptotic giant branch stars with metallicities between $0.04 \leq Z \leq 0.10.$}

\author[Cinquegrana and Karakas]{
Giulia C. Cinquegrana,$^{1,2}$\thanks{E-mail: giulia.cinquegrana1@monash.edu}
Amanda I. Karakas,$^{1,2}$
\\
$^{1}$School of Physics \& Astronomy, Monash University, Clayton VIC 3800, Australia\\
$^{2}$ARC Centre of Excellence for All Sky Astrophysics in 3 Dimensions (ASTRO 3D)\\
}

\date{Accepted XXX. Received YYY; in original form ZZZ}

\pubyear{2021}

\begin{document}
\label{firstpage}
\pagerange{\pageref{firstpage}--\pageref{lastpage}}
\maketitle

\begin{abstract}
\noindent
Low and intermediate mass stars with super solar metallicities comprise a known portion of the universe. Yet yields for asymptotic giant branch (AGB) stars with metallicities greater than $Z=0.04$ do not exist in the literature. This contributes a significant uncertainty to galactic chemical evolution simulations. We present stellar yields of AGB stars for $M=1-8$\(M_\odot\) and $Z=0.04-0.10$. We also weight these yields to represent the chemical contribution of a metal-rich stellar population. We find that as metallicity increases, the efficiency of the mixing episodes (known as the third dredge up) on the thermally pulsing AGB (TP-AGB) decrease significantly. Consequently, much of the nucleosynthesis that occurs on the TP-AGB is not represented on the surface of very metal-rich stars. It instead remains locked inside the white dwarf remnant. The temperatures at the base of the convective envelope also decrease with increasing metallicity. For the intermediate mass models, this results in the occurrence of only partial hydrogen burning at this location, if any burning at all. We also investigate heavy element production via the slow neutron capture process (\textit{s}-process) for three 6\(M_\odot\) models: $Z=0.04, 0.05$ and $0.06$. There is minor production at the first \textit{s}-process peak at strontium, which decreases sharply with increasing metallicity. We find the chemical contributions of our models are dominated by proton capture nucleosynthesis, mixed to the surface during first and second dredge up events. This conclusion is mirrored in our stellar population yields, weighted towards the lower mass regime to reflect the mass distribution within a respective galaxy. \\

\end{abstract}




\section{Introduction}

Stars with super-solar metallicities ($Z\gtrsim2$\(Z_\odot\)) are most commonly located in the central regions and bulges of galaxies, such as the Milky Way, \citep{Lepine11, Feltzing13, Do15} and in giant ellipticals such as M49 \citep{Cohen03}. Metal-rich stars are also found in open clusters such as NGC 6253 \citep{Twarog03}, NGC 6475 \citep{Sestito03} and NGC 6791 \citep{Origlia06}. Here, we define stellar metallicity as the abundance of elements heavier than helium in a star. Irrespective of these known sources, there are only a few studies of very metal-rich stellar models in the literature (see review by \citet{Meynet06} and as discussed in \citet{Karakas21}). There are even fewer studies including AGB stars \citep[e.g.,][]{Bertelli08, Choi16, Marigo17}. In particular, there are no published stellar yields of AGB stars for metallicities greater than about twice solar ($Z\approx0.04$). Stellar yields quantify the mass that is lost from a star over its lifetime and returned to the interstellar medium. They are an essential ingredient in models of galactic chemical evolution \citep[e.g.,][]{Kobayashi20,Romano10,Gibson03,Prantzos18}. The surface abundances used to calculate the yields help us to decipher the mixing and nucleosynthesis processes that occur in stellar interiors. Metal-rich stars are structurally distinct to their low metallicity counterparts \citep{Meynet06} and this significantly impacts upon their chemical contributions to the universe. We will show in this paper that the yields of the most metal-rich stars do not follow a linear progression from those of solar metallicity. Given the galactic mass distribution favours low mass stars, we begin our investigation by looking at the contributions of metal-rich AGB stars. 

In the later phases of their lives, low and intermediate mass stars contribute a significant amount of mass back to the interstellar medium due to strong stellar winds that erode their envelopes. For solar metallicity stars, most of the mass loss occurs when the star is on the TP-AGB phase, which is also when the richest nucleosynthesis occurs. However, in metal-rich stars mass-loss on the early part of the AGB may be so significant, as to remove the envelope before stars even begin thermal pulses (TPs) \citep{Karakas21}. For those that do make it to TP-AGB, we find fewer TPs and mixing events that are less efficient with increasing metallicity. The question that motivates our research is as follows: how does the chemical contribution of low and intermediate mass stars change as we increase the initial stellar metallicity from roughly twice solar ($Z=0.04$) to about 7 times solar ($Z=0.10$), which is the limit of our calculations. In this work, we take solar metallicity to be \(Z_\odot\)$=0.014$ \citep{Asplund09}.

Two major mixing events can occur on the AGB that can significantly alter the chemical composition of the stellar surface. These include the second dredge up (SDU) on the early-AGB and the third dredge up (TDU) during the TP-AGB (see Fig. \ref{fig:HRD}). SDU occurs after the end of core helium burning, when the onset of helium-shell burning causes the outer layers of the star to expand. This causes a decrease in the temperature in this region, sufficient to extinguish the hydrogen burning shell and increase the opacity of the outer layers. For stars with masses greater than $\approx4$\(M_\odot\) (depending on metallicity), the base of the convective envelope can move inwards and mix the products of complete hydrogen burning to the surface. Third dredge up (TDU) occurs in a similar manner, the star expands from the helium-rich intershell region outwards as a result of the instability caused by the geometrically thin helium shell (also known as a TP) \citep{Yoon04}. As opposed to SDU, TDU is usually not a lone event. An AGB star can potentially endure TDU after each TP, which can be tens to thousands of events. The efficiency (and occurrence) of TDU depends heavily on the initial mass and metallicity of the star, with TDU efficiency predicted to increase with decreasing metallicity and/or increasing stellar mass \citep{Karakas02}. 

AGB nucleosynthesis operates in various regions throughout the star. We briefly review the main regions of the star where burning occurs, for further details we refer the reader to \citet{Busso99}, \citet{Herwig05}, \citet{Karakas14} and \citet{Nomoto13}. Most of the nuclear energy is generated via the CNO cycle in the hydrogen burning shell. Ne-Na and Mg-Al chains can initiate at temperatures here of $\approx 1.5\times10^7$K \citep{Mowlavi99sodium}. A smaller amount of energy is generated from the helium burning shell via the triple $\alpha$ reaction, with some energy derived from the $^{12}$C$(\alpha, \gamma)^{16}$O reaction. Further hydrogen burning can occur at the base of the convective envelope in intermediate mass stars ($\approx$ $4-5$\(M_\odot\), but dependent on $Z$). This process is known as hot bottom burning (HBB) and requires higher temperatures to initiate given the lower densities at the base of the envelope. At temperatures of $\approx5\times10^7$K, we see the activation of the CN cycle, where $\approx8\times10^7$K is required for the complete CNO cycle. Temperatures of $\approx 0.8-1.0 \times10^8$K are required during HBB for the Ne-Na and Mg-Al chains to operate here \citep{Forestini97, Ventura13, Karakas03}. 

AGB stars are also a rich source of heavy elements beyond iron through the slow neutron capture process, or \textit{s}-process \citep{Sneden08, Busso99, Kappeler11, Busso04, Gallino98}. In AGB stars, \textit{s}-process elements are produced through a dual combination of neutron sources. The major neutron contributions are provided by the $^{13}$C($\alpha$, n)$^{16}$O reaction during the interpulse phase and lesser contributions from the $^{22}$Ne($\alpha$, n)$^{25}$Mg reaction during TPs. Given that the \textit{s}-process operates in the region between the two shells, efficient TDU is necessary to see the heavy products on the stellar surface. The efficiency of TDU decreases with increasing metallicity \citep{Karakas02, Boothroyd88} and so yields rich in \textit{s}-process elements are favoured in low metallicity stars. \\


There are currently no papers that provide stellar yields from low and intermediate mass stars for metallicities greater than $\sim2$\(Z_\odot\). The metal-rich yields that do exist include the low mass AGB yields ($M=2, 3$\(M_\odot\)) provided by \citet{Battino19} with $Z_{\rm max}=0.03$. \citet{Ventura20} and \citet{Karakas16} produced yields spanning the entire low and intermediate mass range ($M=1-8$\(M_\odot\)) with $Z_{\rm max}=0.04$ ($\approx$ 2\(Z_\odot\)) and $Z_{\rm max}=0.03$. Super-AGB yields are provided by \citet{Siess10} for $M=9,9.5$\(M_\odot\) models, with $Z_{\rm max}=0.04$. There are currently no TP-AGB yields that extend further than $Z=0.04$, although we do find some surface abundances published at higher metallicities. Surface mass fractions of CNO elements along isochrones are provided by \citet{Marigo17}, these extend to metallicities of $Z=0.06$. Final surface C/O ratios of TP-AGB models are provided by \citet{Bertelli08} with $Z_{\rm max}=0.07$ and \citet{Weiss09} with $Z_{\rm max}=0.04$. 


In \citet{Karakas21}, we present new stellar evolutionary sequences of the most metal-rich stars for the range $M=1-8$\(M_\odot\) and $Z=0.04-0.10$. We follow the deviations in the evolutionary tracks of high metallicity stars in comparison to their lower metallicity counterparts. We find distinct variances in the thresholds for nucleosynthetic and mixing processes on the AGB compared to solar metallicity, including the initial mass for core carbon burning. In this paper, we will investigate the consequences of these distinctions on the nucleosynthetic contributions of the most metal-rich stars. We calculate stellar yields for individual stars, as well as collective yields for a metal-rich population. We also present \textit{s}-process yields for $Z=0.04-0.06$ at 6\(M_\odot\). \\


In the next section, \S~\ref{highZ}, we review the specifics of how metallicity affects the structure and processes within the star, as well as the direct consequences for the AGB. In \S~\ref{Methods}, we discuss the methods and input physics involved in our research, specific to the nucleosynthesis calculations. \S~\ref{Results} contains our results, comprised of: surface abundance composition, individual stellar yields, population yields and heavy element yields. Finally, we discuss the implications of these results for the interstellar medium in \S~\ref{Discussion}. 

\section{The impact of a high metallicity on AGB stars}
\label{highZ}

\begin{figure*} 
	\includegraphics[width=18cm]{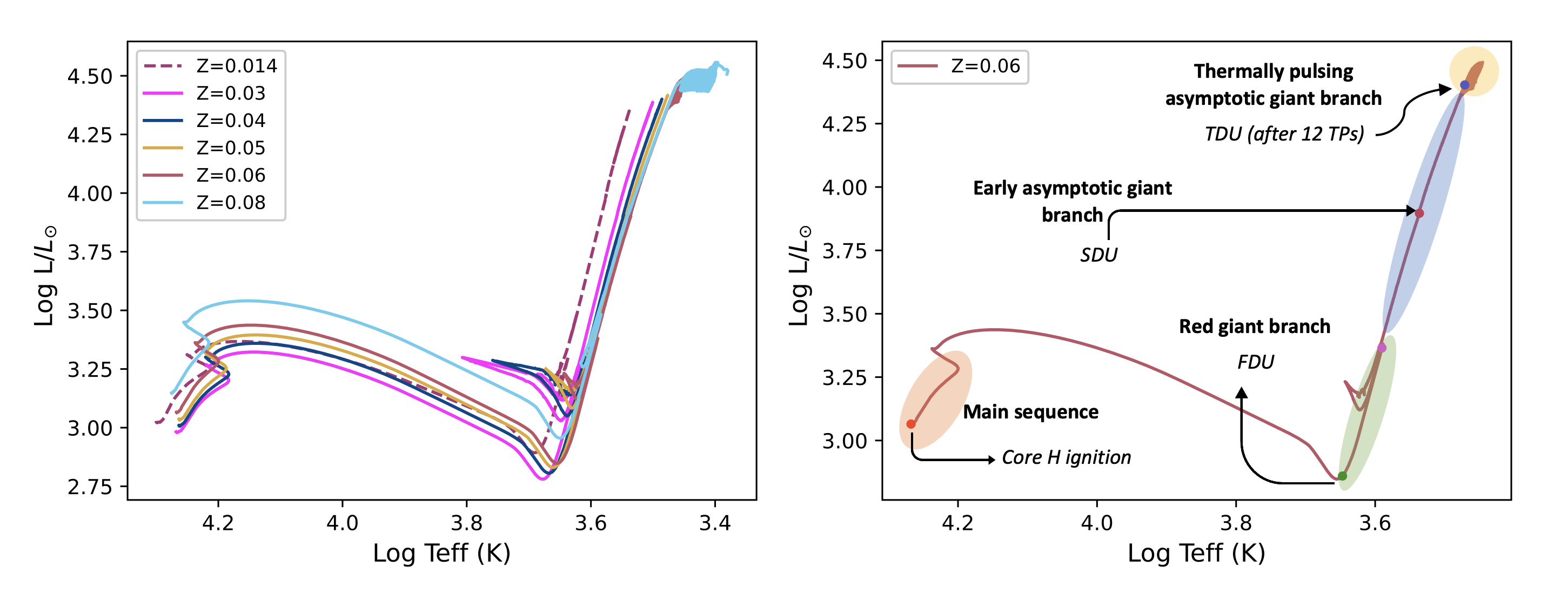}
    \caption{Left panel: Evolutionary tracks of 6\(M_\odot\) models with metallicity varying from $Z=0.014$ to $Z=0.08$. The $Z=0.03$ evolutionary track is from \citet{Karakas14He} and the solar metallicity track from \citet{Karakas16}. Right panel: Isolated track of the 6\(M_\odot\), $Z=0.06$ model with significant phases of evolution and mixing events labelled. The red shaded area designates the main sequence, green the red giant branch, blue the early asymptotic giant branch and yellow the thermally pulsing asymptotic giant branch.}
    \label{fig:HRD}
\end{figure*}

Metallicity has a significant impact on two key stellar variables, opacity ($\kappa$) and mean molecular weight ($\mu$). The effect on these two variables propagates to the nucleosynthesis, mixing processes and mass loss rates of AGB stars \citep{Meynet06, Mowlavi98}. Briefly, as metallicity increases, both $\kappa$ and $\mu$ of the stellar gas increases. A higher $\kappa$ results in both lower luminosities and effective temperatures. This is exhibited in the behaviour of the 6\(M_\odot\), $Z=0.014$ and $Z=0.03$  models in the evolutionary tracks shown in Fig. \ref{fig:HRD}. A higher $\mu$ has the opposite effect: as $\mu$ increases, a gas becomes hotter and more luminous. The effect of the mean molecular weight begins to dominate at metallicities significantly greater than solar. From Fig. \ref{fig:HRD}, we can see that the $Z=0.04$ model experiences a slightly hotter and more luminous main sequence lifetime than $Z=0.03$. The models become progressively hotter as metallicities surpass $Z=0.04$. In fact, the 6\(M_\odot\), $Z=0.08$ model experiences a hotter and more luminous main sequence lifetime than the solar metallicity model. 

A higher metallicity immensely impacts the efficiency of both nucleosynthetic and mixing processes. Due to the lower densities in their envelopes, metal-rich stars have much cooler temperatures at the base of the convective envelope.  For intermediate mass stars, this inhibits the efficiency and occurrence of HBB on the AGB. It also means that a higher initial stellar mass is required for the onset of HBB. In \citet{Karakas21}, we noted that TDU is much less efficient in metal-rich stars; accordingly, we see less nuclear burning products mixed to the surface of the star on the TP-AGB. However, by definition, metal-rich stars will have higher quantities of CNO elements in their initial stellar gas. This results in higher levels of production of some secondary isotopes such as $^{14}$N, which can be mixed to the surface through SDU in intermediate mass stars. The combination of these factors indicates that metal-rich AGB stars will produce significantly different stellar yields than their metal-poor counterparts.

\section{Methods}
\label{Methods}

For this paper we performed a series of nucleosynthesis calculations using the Monash post-processing nucleosynthesis code \citep{Cannon93, Lattanzio96, Lugaro12}. These calculations were executed on stellar evolutionary sequences that are published in \citet{Karakas21}. For details of the methods and input physics involved in running the evolution, we refer the reader to the discussion in \citet{Karakas21} and references therein. Here, we focus on the nucleosynthesis code.

\subsection{Modelling light element nucleosynthesis}

We ran the nucleosynthesis code on the evolutionary sequences calculated by the Monash evolution code which covers the mass and metallicity range $1-8$\(M_\odot\) and $Z=0.04-0.10$. As detailed in \citet{Lugaro12} and \citet{Cannon93}, the nucleosynthesis code takes key structural properties from the evolution code: temperature, density and convective velocities, as a function of time and mass coordinate. It then generates an updated and much more extensive set of abundance changes, due to convective mixing episodes and nuclear reactions, using a larger nuclear reaction network. For the majority of our models, we used a nuclear network of 77 isotope species from hydrogen to sulphur, along with a few iron-peak nuclei \citep{Karakas10Up}. This generates the surface abundance changes from their initial values on the main sequence, through first dredge-up to their final values following the AGB. The initial composition of the elements are scaled solar, with solar abundances provided by \citet{Asplund09}. We use 589 nuclear reaction rates to calculate the nuclear evolution of the models, all rates are from the 2016 default REACLIB database \citep{Cyburt10}. To study the slow neutron capture process in metal-rich models, we also computed a nuclear network of 328 species for the 6\(M_\odot\) models with $0.04\leq Z \leq 0.06$, this is the same nuclear network as described in \citet{Karakas18}. 

\subsection{Modelling heavy element nucleosynthesis}
\label{methods_heavyelements}

AGB stars are an important source of \textit{s}-process elements \citep{Karakas14, Herwig05, Busso99}. In this context, \textit{s}-process elements are produced through two neutron sources. The $^{13}$C($\alpha$, n)$^{16}$O reaction operates during the interpulse phase, at temperatures of approximately $90 \times 10^6$ K \citep{Straniero95, Gallino98}. This reaction releases a steady, low density stream of neutrons over thousands of years \citep{Herwig05}, with densities of $\leq 10^{8}$n cm$^3$ \citep{Fishlock14}. Intermediate mass AGB stars can produce temperatures in the intershell region that exceed $3-3.5 \times 10^8$K during thermal pulses. Where this is reached, the $^{22}$Ne($\alpha$, n)$^{25}$Mg reaction can be activated. Contrary to the $^{13}$C reaction, the $^{22}$Ne source initiates a short ($\approx$years) burst of high neutron density that can approach $10^{12} - 10^{14}$n cm$^3$ \citep{Fishlock14}. 

A significant problem with modelling \textit{s}-process nucleosynthesis in AGB stars concerns the abundance of $^{13}$C and $^{22}$Ne in the intershell. $^{22}$Ne is produced through a process of double $\alpha$ capture on $^{14}$N \citep{Lugaro12}, of which there is sufficient quantity. The general CNO cycle does not produce enough $^{13}$C to be a viable neutron source \citep{Busso99}. Observations however suggest that low-mass stars are the main source of \textit{s}-process elements \citep{Wallerstein98, Busso99}, which means that a way of making $^{13}$C in the intershell is needed. It has been suggested that protons from the envelope can be mixed into the top of the He-shell, making a $^{13}$C pocket. 

The mechanism to generate a pocket of $^{13}$C in the intershell is not fully understood and producing enough $^{13}$C is attempted by various methods in different codes. Some suggested means for the formation of the $^{13}$C pocket include convective boundary mixing \citep[e.g.][]{Herwig00, Cristallo09}, magneto-hydro-dynamical processes \citep[e.g.][]{Trippella16, Busso21} and internal gravity waves \citep{Denissenkov03grav}. We calculate \textit{s}-process nucleosynthesis in 6\(M_\odot\) models, since only intermediate-mass models show TDU at these high metallicities (see \citet{Karakas21}). In our case, we force the formation of $^{13}$C pockets by inserting protons into the top of the helium shell, creating a partially mixed zone in the same manner as \citet{Karakas18} with a mass extent of $1\times10^{-4}$\(M_\odot\). The protons can then be captured by the abundant $^{12}$C nuclei to produce small pockets of $^{13}$C. 

This method allows us to activate the $^{13}$C neutron source, but we do need to consider the realistic nature of proton availability in our chosen models. There will likely be an obtainable source of protons if temperatures are low enough to prevent burning at the base of the envelope during TDU. This is known as hot TDU and has been shown to obstruct \textit{s}-process nucleosynthesis by \citet{Goriely04} and \citet{Herwig04}. Our metal-rich 6\(M_\odot\) models are cooler than lower metallicity intermediate mass stars. At the hottest TP, the $Z=0.04$, 6\(M_\odot\) model reaches a maximum temperature of 4.6 $\times10^7$K at the base of the convective envelope at the maximum extent of dredge up. This is below the required conditions for hydrogen burning at this density, it is consequently reasonable to consider that $^{13}$C pockets can form under these conditions. 

\section{Results}
\label{Results}

\subsection{Surface abundances}
\label{SA} 

\begin{figure*}
	\includegraphics[width=16cm]{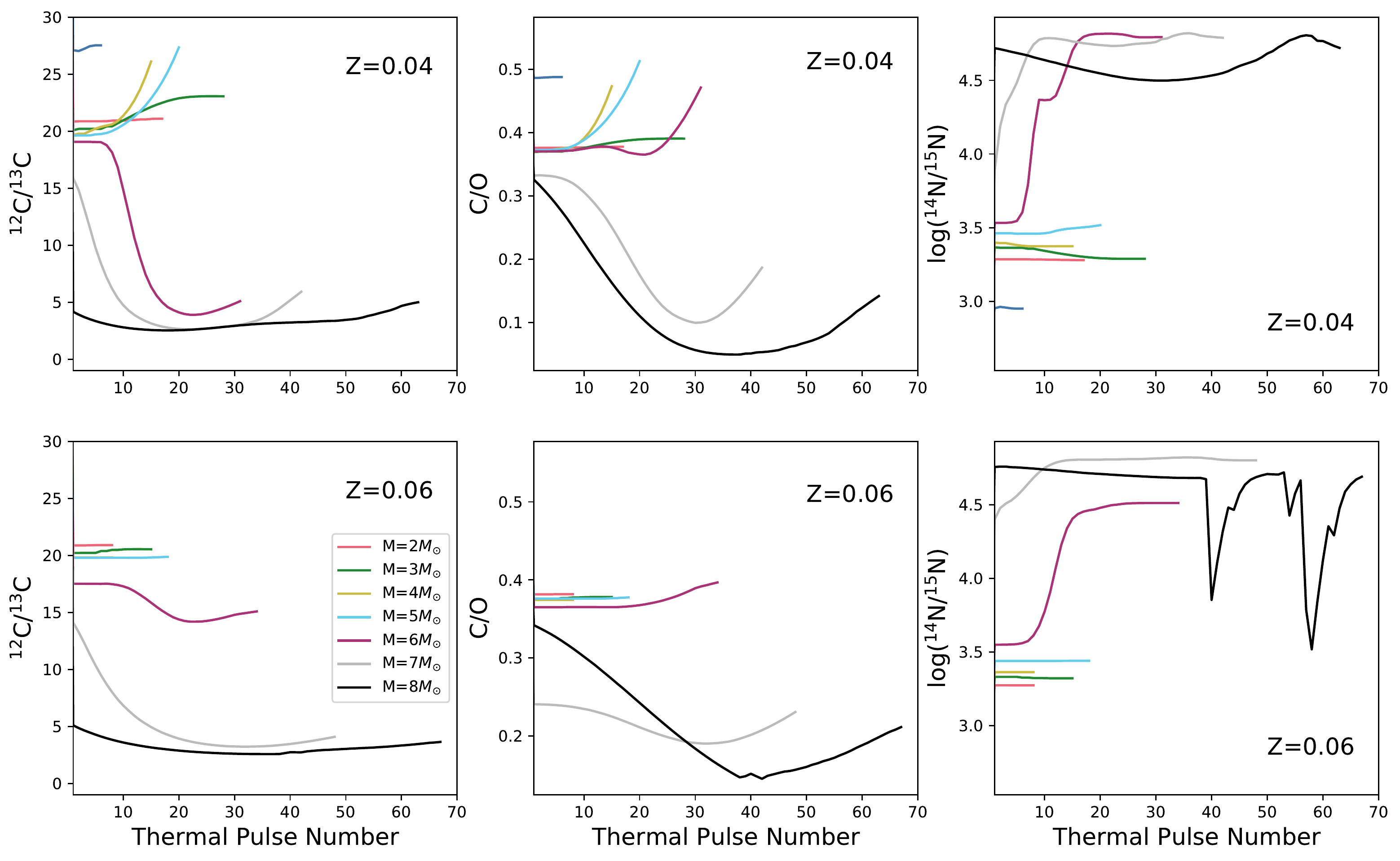}
    \caption{Isotopic ratios $^{12}$C/$^{13}$C, C/O and $^{14}$N/$^{15}$N for $Z=0.04$ (top) and $Z=0.06$ (bottom). The first column shows $^{12}$C/$^{13}$C, where horizontal lines indicate that models are not massive enough to endure TDU. A positive gradient indicates the onset of TDU and a negative gradient indicates the onset of HBB. The second column, C/O, shows the same indications for TDU and HBB as previously mentioned. The third column plots log10($^{14}$N/$^{15}$N) on the y axis, the onset of HBB results in a large increase in $^{14}$N.}
    \label{fig:ratios_all}
\end{figure*}

In this section, we focus on the surface composition of our models during the TP-AGB. We first provide a summary of the main features of the metal-rich models presented in \citet{Karakas21}, in terms of the mixing experienced on the AGB.

As metallicity increases, the mass threshold for TDU and HBB also increases. For example, TDU begins at 2\(M_\odot\) at solar metallicity, 5\(M_\odot\) for $Z=0.06$ and 7\(M_\odot\) for $Z=0.08$. As discussed in our first paper, we find no TDU for models with $Z\geq 0.09$. HBB initiates at 4.5\(M_\odot\) at solar metallicity, this is delayed until 7\(M_\odot\) at $Z=0.10$. Additionally, the efficiency for both processes decreases with increasing metallicity. We will discuss the implications of this in more detail in \S~ \ref{Discussion}, given their importance for nucleosynthesis. We refer the reader to \citet{Karakas21} for further information regarding our models, where we have tabulated some important features of each model (occurrence of SDU, number of TPs, maximum luminosities and temperatures, lifetimes) for each metallicity. Prior to divulging the results of our stellar yields, we first review the isotopic ratios $^{12}$C/$^{13}$C, C/O and $^{14}$N/$^{15}$N at the stellar surface. These ratios are commonly used in AGB modelling to indicate the quantity, duration and efficiency of TDU and HBB. In the context of this paper, we are most interested in how these ratios change as a function of metallicity. 

Fig. \ref{fig:ratios_all} shows the $^{12}$C/$^{13}$C, C/O and $^{14}$N/$^{15}$N ratios as functions of TP number for $Z=0.04$ and $0.06$ (where TP is a proxy for time on the TP-AGB). A PDF containing the rest of the metallicity range is available as part of our online material. The $^{12}$C/$^{13}$C ratio comprises the first column of Fig. \ref{fig:ratios_all}, where the low mass models that are not capable of TDU produce a horizontal line. As TDU ensues, the $^{12}$C produced in the helium burning shell is mixed to the surface, this increases the $^{12}$C/$^{13}$C ratio. This occurs at 3.5\(M_\odot\) for $Z=0.04$ and 5\(M_\odot\) for $Z=0.06$. With the onset of HBB (5.5\(M_\odot\) for $Z=0.04$ and 6\(M_\odot\) for $Z=0.06$), $^{12}$C is burnt into $^{14}$N, lowering the $^{12}$C/$^{13}$C ratio. Similar designations apply to the C/O ratio in the second column of Fig. \ref{fig:ratios_all}. Given that the surface abundances of $^{14}$N and $^{15}$N are unaffected by TDU, the $^{14}$N/$^{15}$N ratio remains almost constant for low mass stars. With the onset of HBB, the ratio increases with the subsequent production of $^{14}$N. 

No TDU episodes occur at $Z\geq 0.09$, for the entire low and intermediate mass range. As a result, we predict that there will be very little positive movement in both the $^{12}$C/$^{13}$C and C/O ratio plots as we reach the highest metallicities. Even from our lowest metallicity, $Z=0.04$, carbon star formation is completely prevented (defined where C/O $\geq 1$) \citep{Wallerstein98, Karakas14He}. However, even with the lower temperatures at the base of the convective envelope, there is still significant indication of HBB in the $Z=0.06$ models. The $Z=0.1$ models (Fig. \ref{fig:ratios_all_A_all}) behave vastly different to the lower metallicities plotted here. Within the $1-8$\(M_\odot\) mass range, only three models enter the thermally pulsing AGB phase: 5, 5.5 and 6\(M_\odot\). Of the $Z=0.10$ models that endure thermal pulses, there is no TDU. Mild HBB is found in the 7 and 7.5\(M_\odot\) models, occurring on the early-AGB. 
The inability of the $M>6$\(M_\odot\) models to endure thermal pulses is due to extreme mass loss at the maximum metallicity range. We show in our first paper that a 6.5\(M_\odot\), $Z=0.10$ model without mass loss can enter the TP-AGB, with 37 thermal pulses. This model (with no mass loss) also endures HBB, but shows no evidence of any TDU. 

\subsection{Stellar yields}
\label{stellaryields}

Stellar yields represent the integrated mass that is expelled from a star over its lifetime. In other terms, they represent the contribution of a star to the interstellar medium. The stellar yields are shaped by the same processes discussed in section \ref{SA}, along with mass-loss which is responsible for ejecting the outer layers back to the interstellar medium. Using stellar models, we can predict the chemical output of stars of various metallicities and masses in a population. By doing so, we aim to match elemental production (and destruction) within a respective galaxy with the responsible stellar mechanisms. Quantitatively, \citet{Karakas07} define the yield of a species, \textit{i}, as: 

\begin{equation}
M_i = \int_{0}^{\tau}[X(i)-X_0(i)]\frac{dM}{dt}dt. 
\label{Yield}
\end{equation}

\noindent
The yield of species \textit{i} is denoted by $M_i$ (in \(M_\odot\)), $\frac{dM}{dt}$ represents the current mass loss rate at time $t$, the species current and initial mass fractions at the surface are given by $X(i)$ and $X_{0}(i)$ and $\tau$ is the lifetime of the model. A negative yield reflects the quantity of species \textit{i} that is destroyed inside the star over its lifetime (compared to the initial amount present in the wind). Conversely, a positive yield quantifies the amount of species \textit{i} produced over the star's lifetime. For low and intermediate mass stars, stellar yield contributions occur mainly on the giant branches (especially the AGB) given that this is where stellar winds begin to significantly erode their extended envelopes. In the following sub sections, we will follow 17 isotopes that are known to be affected on the AGB and examine how their yields are altered due to increasing metallicity. We have included solar metallicity models from \citet{Karakas16} in the following figures for comparison. Tabulated yields are available as part of our online material.

\begin{figure*}
	\includegraphics[width=16cm]{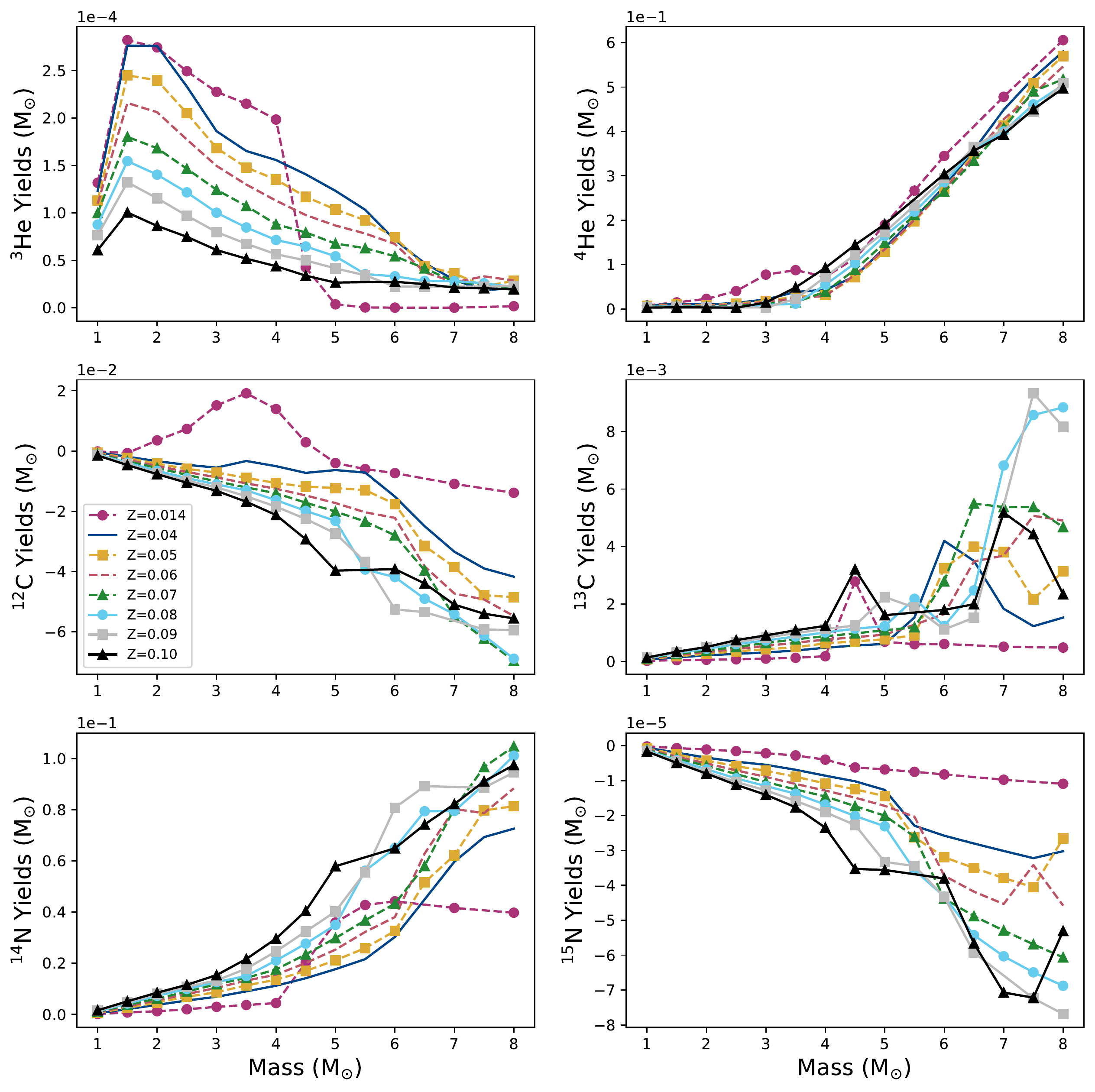}
    \caption{Top panel (a): $^{3}$He yields. The highest production of this isotope is found in low mass stars. $^{3}$He decreases with metallicity, given that as the Z mass fraction increases, the hydrogen mass fraction decreases. There is a steep drop-off in the solar metallicity model at 4.5\(M_\odot\) where $^{3}$He is destroyed in HBB. Top panel (b): $^{4}$He yields. $^{4}$He is less dependent on metallicity. Given that $^{4}$He is the main product of hydrogen burning, we see a steep increase in the $^{4}$He yields with both SDU and the onset of HBB. Middle panel (a): $^{12}$C yields. Production increases with TDU in the $Z=0.014$ model, until the onset of HBB where it is destroyed. For $Z>0.014$, the isotope is solely destroyed. Middle panel (b): $^{13}$C yields. The production of this isotope occurs mainly in FDU and mild HBB. Bottom panel (a): $^{14}$N yields. The highest levels of $^{14}$N are created in stars capable of both SDU and HBB (i.e. those greater than 4\(M_\odot\)). For the higher metallicity models, this is likely a combination of both primary and secondary nitrogen production. Bottom panel (b):  $^{15}$N yields. There is a net destruction of this isotope for all metallicities. This destruction increases with both high metallicity and the onset of HBB.}
    \label{fig:nonweighted_1}
\end{figure*}

\subsubsection{Helium}

$^{4}$He is the main product of hydrogen burning, consequently both first dredge up (FDU) on the red giant branch and second dredge up (SDU) on the early AGB cause an increase in the surface $^{4}$He abundance \citep{Becker79}. Fig. \ref{fig:nonweighted_1} shows a steep increase in $^{4}$He at 4\(M_\odot\), which corresponds to the approximate mass threshold for SDU. Further contributions will be made by intermediate mass stars on the AGB with temperatures sufficient for HBB. However, for the cases of $Z=0.09$ and $0.10$ which undergo very little HBB, their abundances are primarily the result of FDU and SDU for the intermediate mass models. From \citet{Karakas21}, we found SDU to be relatively metallicity invariant in intermediate mass models. 

Low-mass stars ($M\leq2$\(M_\odot\)) are the main sources of $^{3}$He. This begins pre main sequence, where initial deuterium is converted to $^{3}$He, then further built up by p-p reactions on the main sequence. We note that in our models, we assume zero $^{3}$He abundance initially. Thus, all production of this isotope in our models stems from hydrogen burning on the main sequence which is mixed to the stellar surface with FDU \citep{Iben67, Iben67stellar, Charbonnel98}. The $^{3}$He yield is indicative of the depth of FDU, as well as the burning conditions during the main sequence. From Fig. \ref{fig:nonweighted_1}, we see a peak in this yield for all metallicities at $\approx$ 1.5\(M_\odot\). From \citet{Karakas21}, the true maximum depth of FDU is between $2-2.5$\(M_\odot\). For the solar metallicity model, we see a sharp decline in Fig. \ref{fig:nonweighted_1} with the onset of HBB between 4 and 4.5\(M_\odot\). $^{3}$He is destroyed during this process to produce $^{7}$Li, due to p-p reactions that occur at the hot base of a well-mixed envelope \citep{Smith89Li, Smith90Li, Travaglio01Lith}. 

\subsubsection{Carbon}

AGB stars produce $^{12}$C via the triple $\alpha$ process in the helium burning shell. Consequently, surface changes in $^{12}$C occur with the onset of TDU. The greatest surface enrichment of this isotope then correlates to the models that endure the largest number of TDU episodes (with temperatures low enough to avoid destruction by HBB) \citep{Marigo01, Ventura13}. At solar metallicity, this occurs in models with initial masses between $2-4.5$\(M_\odot\) \citep{Karakas14He}. From Fig. \ref{fig:nonweighted_1}, maximum $^{12}$C production at $Z=0.014$ in our models occurs at an initial mass of approximately 3.5\(M_\odot\). Beyond 3.5\(M_\odot\) for the same metallicity, the $^{12}$C yield negatively trends with the onset of HBB where $^{12}$C is burnt into $^{13}$C and $^{14}$N during the interpulse phase \citep{Lattanzio03, Ventura13, Ventura18}. Fig. \ref{fig:nonweighted_1} demonstrates that as metallicity increases above $Z=0.014$, there is little to no production of $^{12}$C. In fact, the destruction of this isotope increases with increasing initial mass. This is a consequence of the inefficient TDU that correlates with increasing metallicity.

The $^{13}$C yields experience an initial increase as a result of FDU, with further contributions from mild HBB. The peaks of $0.04\leq Z<0.08$ correspond to the onset of HBB. We also see smaller peaks for lower initial masses at $0.08\leq Z\leq 0.10$. These correlate to the mass range for the onset of SDU. The greatest temperatures at the base of the convective envelope (by proxy, the most efficient HBB) occur in the solar metallicity model. Yet regardless of this, the $Z=0.014$ models exhibited in Fig. \ref{fig:nonweighted_1} show the lowest production of $^{13}$C. This is indicative that the higher abundances of $^{13}$C as Z increases are a mix of both secondary $^{13}$C, mixed to the surface during FDU (and SDU in intermediate mass stars) and some primary $^{13}$C produced during HBB on the AGB. 

\subsubsection{Nitrogen}

From Fig. \ref{fig:nonweighted_1}, the first significant increase in $^{14}$N (for all metallicities) occurs around the $\approx$ 4\(M_\odot\) threshold with the onset of SDU. There is a subsequent increase in production of $^{14}$N in models capable of enduring HBB on the AGB, this boundary is $\approx$ 4.5\(M_\odot\) at $Z=0.014$ and $6-7$\(M_\odot\) for $Z=0.04-0.10$. Due to the anti-correlation of metallicity and the temperatures at the base of the convective envelope (i.e. low metallicity stars endure the full CNO cycle during HBB at lower initial masses), low metallicity stars are generally more efficient producers of $^{14}$N \citep{Ventura13}. However, our models actually show that greater quantities of $^{14}$N are produced as metallicity increases. This is due to a greater contribution of secondary $^{14}$N from the higher metallicity models. Given the conservation of ingredients when TDU is inefficient, metal-rich gas will produce more $^{14}$N prior to the AGB than a lower metallicity gas (and destroy more $^{12}$C and $^{16}$O, $^{18}$O). These products are then mixed to the surface during FDU and SDU in intermediate mass stars. In the models without efficient TDU and HBB, the dominant mode of nitrogen production will be secondary. In the models that experience some TDU and HBB, there will be a mix of some primary production of $^{14}$N at the expense of $^{12}$C (mixed up from the helium burning shell), alongside some secondary $^{14}$N.

There is a net destruction of $^{15}$N for all metallicities exhibited in Fig. \ref{fig:nonweighted_1}. The destruction of $^{15}$N in stars with initial masses below $3-4$\(M_\odot\) is due to hydrogen burning on the main sequence, where the surface is diluted during FDU \citep{Marigo01}. Stars with initial masses greater than 4\(M_\odot\) can experience further depletion from both HBB and SDU. At $5-6$\(M_\odot\), the destruction rates deepen with increasing metallicity. The greater initial CNO content of the very metal-rich gases allows for greater destruction of $^{15}$N during HBB and SDU. 

\begin{figure*}
	\includegraphics[width=16cm]{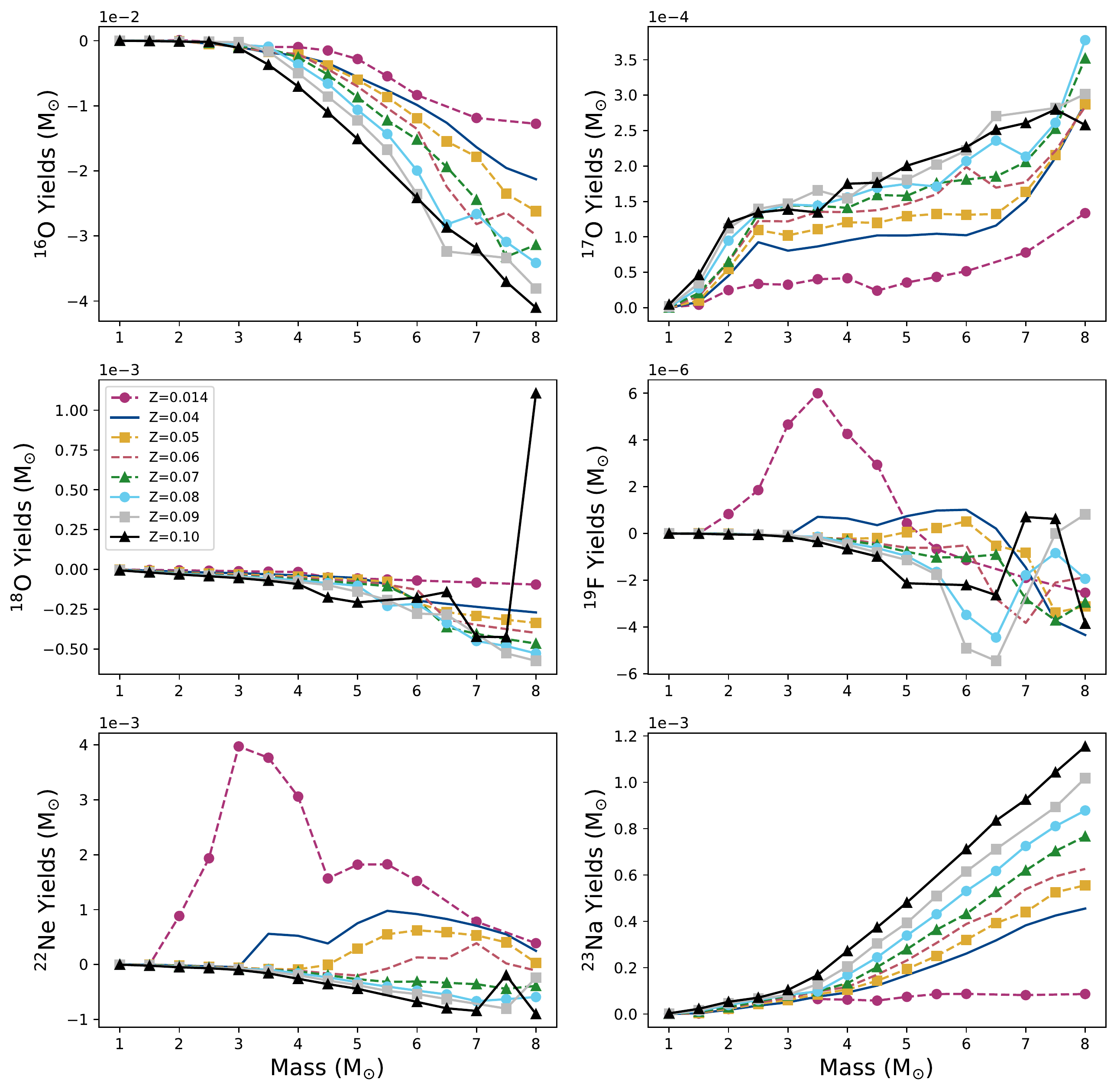}
    \caption{Top panel (a): $^{16}$O yields. This isotope is formed primarily during $\alpha$ capture and is thus dependent on TDU for this to be reflected on the stellar surface. All models show consistent destruction. Top panel (b): $^{17}$O yields. There is some production during hydrogen burning which increases with metallicity. This is due to the greater quantity of CNO seeds in their initial gas. Middle panel (a): $^{18}$O yields. For the same reason, destruction increases with metallicity for both $^{16}$O and $^{18}$O. Middle panel (b): $^{19}$F yields. Given that $^{19}$F is predominantly made in the helium intershell region, TDU is needed to mix this isotope to the surface. Hence, its peak value will correspond to the mass range with the most efficient TDU. We see this only reflected in the $Z=0.014$ model.  The main destruction of $^{19}$F occurs as a result of HBB. Bottom panel (a): $^{22}$Ne yields. $^{22}$Ne is produced via $\alpha$ capture in the helium rich intershell region and is mixed to the surface with TDU. The highest production of this isotope will be in the low mass, low metallicity range. $^{22}$Ne is destroyed with the onset of HBB. Bottom panel (b): $^{23}$Na yields. We expect to see $^{23}$Na production in models that experience strong HBB (with temperatures sufficient to activate the Ne-Na chain). In the $Z=0.014$ there is relatively no production. The higher metallicity models all show relatively strong sodium production, increasing with Z. This implies that the $^{23}$Na is secondary and thus depends on the initial CNO quantity in the gas.}
    \label{fig:nonweighted_2}
\end{figure*}

\subsubsection{Oxygen}

Oxygen is composed of three stable isotopes: $^{16}$O, $^{17}$O and $^{18}$O. The dominant isotope, $^{16}$O, is primarily formed via $\alpha$ capture in the reaction $^{12}$C($\alpha,\gamma)^{16}$O. Given the composition of the intershell region is heavily dominated by $^{4}$He (rather than the necessary $^{12}$C) and combined with the short timescale of TPs, any production of $^{16}$O is relatively minor \citep{Karakas14}. As a consequence, surface abundance changes of $^{16}$O increase marginally with the operation of TDU. Peak $^{16}$O abundance in AGB stars is subsequently associated with low mass, low metallicity models that do not endure HBB (given that $^{16}$O is destroyed during the ON cycle to produce $^{14}$N) and experience a greater number of efficient TDU episodes \citep{Herwig04, Herwig2004dredge, Karakas10Up}. For the $Z=0.10$ models that do not undergo HBB, destruction of this isotope occurs prior to the AGB during hydrogen burning on the main sequence and is reflected on the surface as a result of SDU. 

There is some production of $^{17}$O during hydrogen burning, conditional upon the reactions
$^{16}$O($p,\gamma)^{17}$F($\beta^+ \nu)^{17}$O dominating over the destruction channels $^{17}$O($p,\gamma)^{18}$F and $^{17}$O($p,\alpha)^{14}$N \citep{Marigo01}. As discussed previously within the context of $^{14}$N, high metallicity models begin with greater initial quantities of CNO elements. Thus as shown in Fig. \ref{fig:nonweighted_2}, the isotopes $^{16}$O and $^{18}$O are more readily destroyed for high metallicity. By the same notion, larger quantities of $^{17}$O are produced simultaneously.

Given that SDU efficiency increases with initial stellar mass, the models at the top end of the mass range endure deep SDU (see Fig. 5 from \citet{Karakas21}). This results in significant amounts of $^{18}$O dredged into the envelope, which can then be burned to $^{19}$F with the onset of HBB. We note that the unusually high $^{18}$O yield for $Z=0.10$ is a result of the evolution code cutting off prior to the occurrence of any HBB. We speculate that if the evolution were to be continued that the $^{18}$O would be destroyed in a similar fashion to the rest of the intermediate mass range. 

\subsubsection{Fluorine}

AGB stars are a recognized production site for $^{19}$F in the Galaxy \citep{Lugaro04reaction, Cristallo09, Jorissen92}. In AGB stars, $^{19}$F is produced during the convective thermal pulses as well as during the interpulse periods, in the top layer of the helium intershell region. As detailed in \citet{Lattanzio99}, $^{18}$O requires protons produced via neutron capture on $^{14}$N through the $^{14}$N($n,p$)$^{14}$C reaction. If the $^{18}$O can capture these protons (potentially also from $^{26}$Al($n,p$)$^{26}$Mg), then the following series can occur: $^{18}$O($p,\alpha$)$^{15}$N($\alpha,\gamma$)$^{19}$F.

We note that the work of \citet{Abia19} and \citet{Vescovi21} have shown recently that the contributions of $^{19}$F from AGB stars are a significantly smaller percentage of the budget than originally predicted by \citet{Jorissen92}. \citet{Vescovi20magnetic, Vescovi21} incorporate \textit{magnetically induced} $^{13}$C pockets in their models which result in lower proton abundances. They thus produce less $^{14}$N and $^{19}$F than models with $^{13}$C pockets produced by convective overshoot. However, the discussion of overproduction of $^{19}$F applies to models that contain a $^{13}$C pocket, of which the models discussed in this section do not. Our models that \textit{do} contain pockets are of intermediate mass, and so are not significant producers of this isotope.

$^{19}$F production in AGB stars is heavily dependent on TDU efficiency \citep{Lugaro04reaction}, given that it is needed to mix the $^{19}$F produced in the intershell into the envelope. From Fig. \ref{fig:nonweighted_2}, we can see that $^{19}$F production peaks with the highest TDU efficiency, which correlates to the solar metallicity 3.5\(M_\odot\) model. There are also relative increases in abundances for $Z=0.04, 0.05$ and $0.06$ when TDU begins. A downwards trend is observed into the intermediate mass range. The relative abundance of $^{19}$F that has been successfully mixed into the envelope will decrease with the onset of HBB, as a result of the main destruction channel $^{19}$F($p,\alpha$)$^{16}$O \citep{LaCognata11}.  

\begin{figure} 
	\includegraphics[width=0.9\columnwidth]{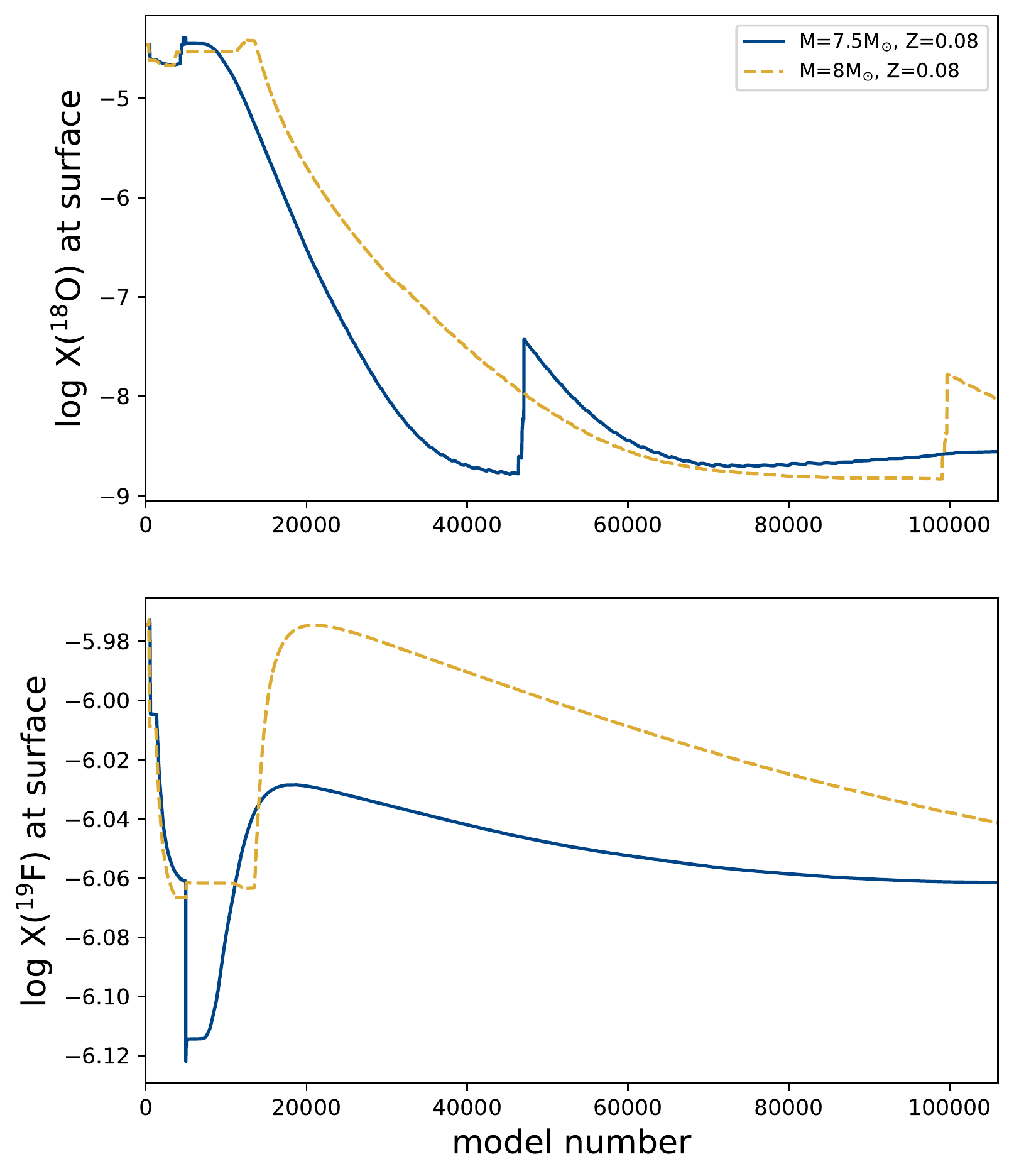}
    \caption{Surface abundances (in mass fraction) of $^{18}$O and $^{19}$F as a function of model number, where model number serves as a proxy for time. A deep SDU mixes a significant quantity of $^{18}$O in the envelope of these models. As temperatures at the base of the convective envelope increase, $^{19}$F is produced at the expense of $^{18}$O. Once the concentration of $^{18}$O is significantly depleted and temperatures surpass 7$\times 10^7$K, the $^{19}$F is then destroyed at a rate proportional to the maximum temperatures reached.}
    \label{fig:oxy_fluorine_z08}
\end{figure}

Some interesting behaviour can be seen at the highest metallicities. For the models of $M>7$\(M_\odot\) and $Z\geq0.08$, we see an increase in the yields of $^{19}$F. The increase is caused by a deep SDU (see Fig. 5 in \citet{Karakas21}). The convective envelope is able to reach into the top of the helium exhausted shell and dredge up a significant quantity of $^{18}$O. In Fig. \ref{fig:oxy_fluorine_z08}, we have plotted the surface mass fractions of $^{18}$O and $^{19}$F as a function of model number (where model number is a proxy for time). We use the 7.5 and 8\(M_\odot\), $Z=0.08$ models as examples. For both masses, $^{18}$O is brought into the envelope during SDU. From model numbers 6,300 (7.5\(M_\odot\)) and 13,400 (8\(M_\odot\)), temperatures at the base of the envelope exceed 5$\times 10^7$K. From there, the $^{18}$O(p, $\gamma$)$^{19}$F reaction can ensue. As the $^{18}$O concentration runs out and temperatures increase, $^{19}$F begins to be destroyed via the destruction channel mentioned above, $^{19}$F($p,\alpha$)$^{16}$O. The lower temperatures of our metal-rich models minimise the destruction of $^{19}$F once it has been produced. As temperatures increase significantly, as in the case of the 8\(M_\odot\) model, this isotope is quickly destroyed.

\subsubsection{Neon and sodium} 

In this section we focus on only one of the three stable neon isotopes, $^{22}$Ne. This is the main neon isotope that we predict is produced in AGB stars, as suggested by elemental neon enhancements in planetary nebulae (PNe) \citep{Karakas03neon}. Evidence for substantial $^{22}$Ne production is observed in silicate carbide (SiC) grains. SiC grains condense in the carbon-rich environment surrounding low mass AGB stars and trap noble gases such as Neon, providing a strong constraint on theoretical nucleosynthesis models \citep{Gallino90SiC, Anders93SiC, Lugaro03SiC, Heck07SiC}. 

In AGB stars, the intershell region between the inner helium burning shell and the outer hydrogen burning shell is enriched in $^{14}$N. This $^{14}$N can undergo $\alpha$ captures to form $^{22}$Ne. If temperatures in the intershell region exceed $3-3.5 \times 10^8$K, the $^{22}$Ne can be subject to further $\alpha$ captures to form $^{25}$Mg and $^{26}$Mg \citep{Denissenkov03, Karakas06}. Otherwise, with the next TP, the $^{22}$Ne is dredged up into the convective envelope. In the next interpulse period, the Ne-Na chain can be activated in both the hydrogen burning shell and at the base of the convective envelope. 

A maximum $^{22}$Ne yield will occur in models with the most efficient dredge up episodes \citep{Cristallo15, Karakas10Up}, but that are not hot enough at the base of the envelope to initiate the Ne-Na chain. \citet{Karakas03neon} found that peak intershell abundances of $^{22}$Ne occur at 3\(M_\odot\) for their metallicity range ($Z=0.004-0.02$). This is reflected in the solar metallicity model in Fig. \ref{fig:nonweighted_2}. For our metallicity range, this peak shifts to higher initial masses, given that efficient TDU is a requirement. There is no substantial increase in $^{22}$Ne for $Z>0.06$, due to very inefficient TDU ($Z=0.07, 0.08$) or a complete lack of TDU ($Z=0.09$ and $0.10$). Temperatures high enough to activate the Ne-Na chain are only achieved for the $7-8$\(M_\odot\) range at $Z=0.04$ and 8\(M_\odot\) for $0.04<Z\leq0.09$. Thus our $^{22}$Ne yields are more largely a factor of inefficient dredge up rather than destruction via the Ne-Na chain. 

The $^{23}$Na yields in Fig. \ref{fig:nonweighted_2} show a relative plateau until about $3-4$\(M_\odot\). With the onset of HBB, there is some competition between the production of $^{23}$Na via the Ne-Na chain and the two main destruction channels: $^{23}$Na($p,\alpha$)$^{20}$Ne and $^{23}$Na($p,\gamma$)$^{24}$Mg. These reactions depend heavily on the temperature in the envelope and the extent of the burning that occurs \citep{Hale01, Vaughan18, Slemer16}. Although the intermediate mass models endure HBB, none reach temperatures capable of activating the Ne-Na chain until $M\geq7$\(M_\odot\). However, the $^{23}$Na surface abundance also increases during SDU, which occurs consistently with the increase in sodium for our range. 

It is interesting to note here that there is very little production of $^{23}$Na in the solar metallicity models in comparison to the metal-rich models presented in Fig. \ref{fig:nonweighted_2}. In fact, the yield of $^{23}$Na scales with increasing metallicity. This implies that $^{23}$Na is secondary in AGB models and strongly dependent on the initial seeds of CNO present in the gas. This is likely to be different to production at lower metallicities where substantial $^{12}$C production occurs, see for example \citet{Cristallo06}.

\begin{figure*}
	\includegraphics[width=16cm]{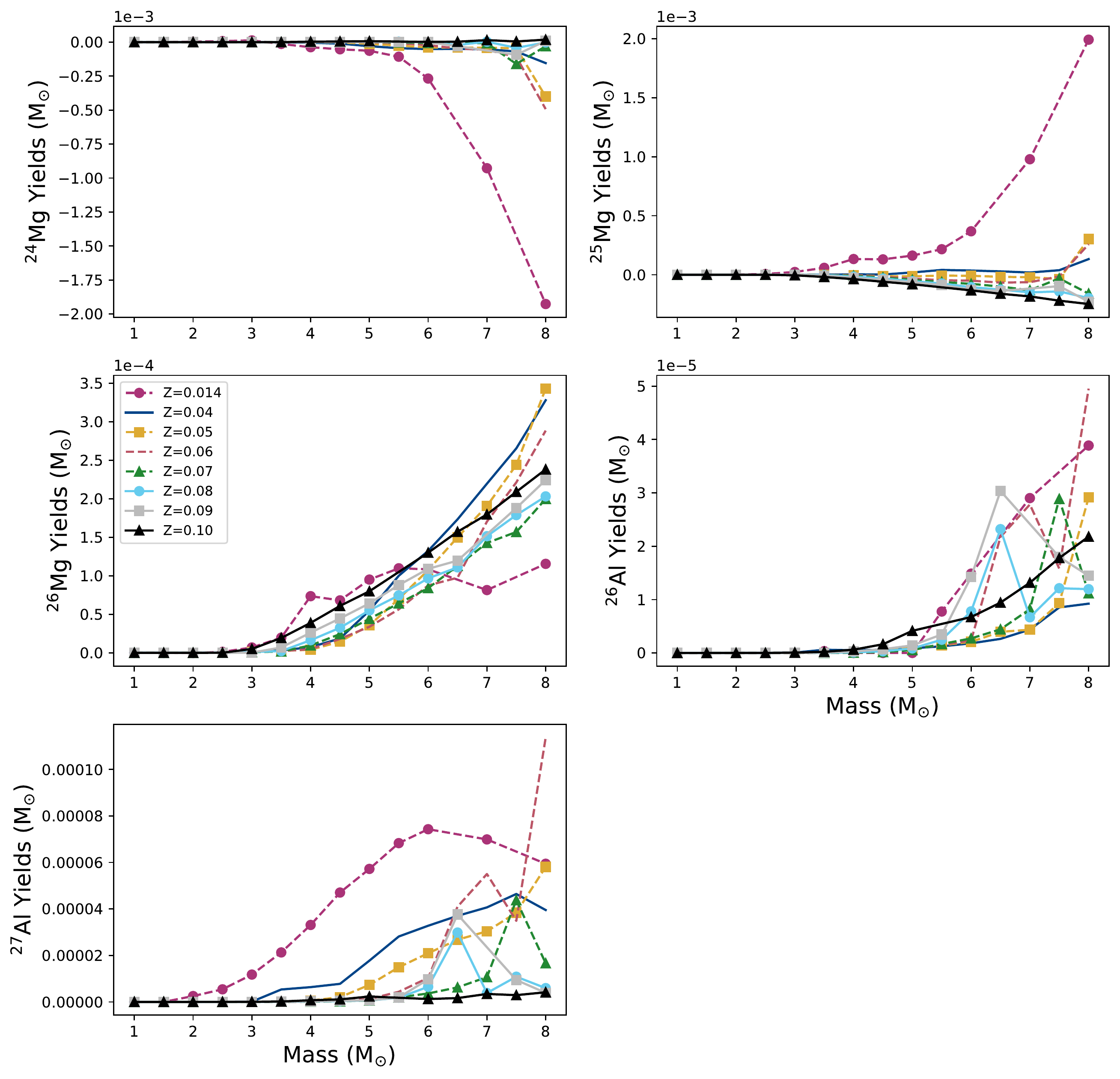}
    \caption{Top panel (a): $^{24}$Mg yields. AGB stars are not substantial producers of $^{24}$Mg, variations in abundance occur only for those that reach temperatures capable of burning $^{24}$Mg to produce aluminium. In this case, this occurs only in the solar metallicity models. Top panel (b): $^{25}$Mg yields. $^{25}$Mg is produced in substantial quantities in the helium burning shell. Thus, changes in the surface abundance of this isotope require TDU for it to be reflected on the surface. This favours low metallicity stars. Middle panel (a): $^{26}$Mg yields. The production of $^{26}$Mg increases with both mass and metallicity, except in the $Z=0.014$ model. Middle panel (b): $^{26}$Al yields. The highest production of $^{26}$Al ensues in models that reach temperatures sufficient for HBB. It is therefore expected that lower metallicity models will experience the highest production. In this case however, the $Z=0.10$ model has the second highest production of this isotope following $Z=0.014$, likely a result of SDU. Bottom panel (a): $^{27}$Al yields. For $Z=0.014$, we see some evidence of $^{27}$Al produced in the hydrogen burning shell and mixed to the surface with TDU. With onset of HBB, production increases significantly. There is some production of this isotope in the higher metallicity models that undergo HBB.}
    \label{fig:nonweighted_3}
\end{figure*}

\subsubsection{Magnesium and aluminium}

Intermediate mass AGB stars are one of the main production sites of $^{25}$Mg and $^{26}$Mg \citep{Karakas03} as well as $^{26}$Al \citep{Forestini91, Arnould87, Norgaard80, Frantsman89}. The ratio of magnesium and aluminium isotopes are re-arranged via the Mg-Al chain. Analogous to the requirements of the Ne-Na chain, the Mg-Al can occur in AGB stars in both the hydrogen burning shell and at the base of the convective envelope. $^{25}$Mg and $^{26}$Mg can also be produced in a third site, in the helium burning shell. $\alpha$ captures on $^{22}$Ne can produce significant amounts of these isotopes. There is no aluminium production in the helium burning shell, however depletion of $^{26}$Al can occur via the $^{26}$Al($n,p$)$^{26}$Mg and $^{26}$Al($n,\alpha$)$^{23}$Na reactions \citep{Karakas03}. 

Fig. \ref{fig:nonweighted_3} shows substantial destruction of $^{24}$Mg at solar metallicity and lesser destruction for $0.04\leq Z\leq0.06$ at 8\(M_\odot\). There is very little change in $^{24}$Mg for models with $Z>0.06$, given these high metallicity models do not reach temperatures sufficient for significant $^{24}$Mg depletion via the Mg-Al chain. The $Z=0.014$ models also experience substantial $^{25}$Mg production in Fig. \ref{fig:nonweighted_3} as a result of efficient TDU mixing this isotope from the helium burning shell to the surface. Again, there is some production of $^{25}$Mg at 8\(M_\odot\) for $Z=0.04-0.06$. Modest destruction occurs for the rest of the range, which increases with metallicity. There is a uniform increase in $^{26}$Mg for all models, except for $Z=0.014$ which tapers off once it reaches a maximum at 5.5\(M_\odot\). 

Given that the production of $^{26}$Al and $^{27}$Al requires high temperatures both in the hydrogen burning shell and at the base of the convective envelope, it follows that the lower metallicity models will be richer in these isotopes. Fig. \ref{fig:nonweighted_3} shows that there is relatively no production of $^{26}$Al before the onset of HBB. For $^{27}$Al, production gradually increases in the $Z=0.014$ model starting at 2\(M_\odot\), indicating that there is some $^{27}$Al production in the hydrogen burning shell that is mixed to the surface with TDU. For the rest of the metallicity range, we see a tapered production of increasing $^{27}$Al with decreasing metallicity.

\subsection{Comparison of yields with other studies}

\begin{figure} 
	\includegraphics[width=0.9\columnwidth]{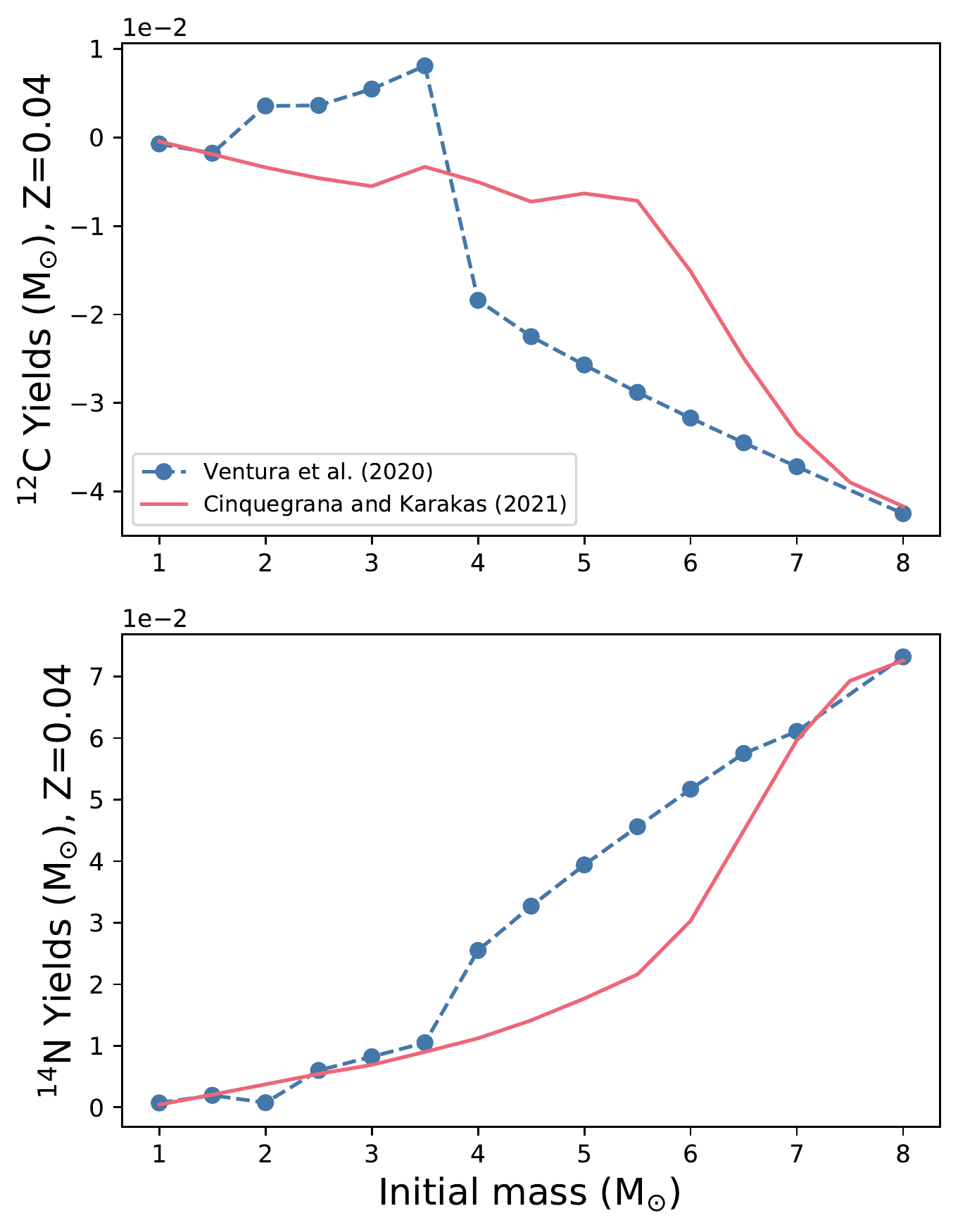}
    \caption{Comparison of the $^{12}$C and $^{14}$N yields of \citet{Ventura20} and those in this work for $Z=0.04$.}
    \label{fig:yield_comparison}
\end{figure}

The only published yields in the literature that fall within our mass and metallicity range are those of \citet{Ventura20} (hereafter, V20). We confirm the comparison they make to the yields published in \citet{Karakas16} based on the $Z=0.03$ models from \citet{Karakas14He}. We update this to our current nucleosynthesis yields at $Z=0.04$. In the low mass range ($1-3$\(M_\odot\)), we see very similar yields between the two codes. Particularly for $^{13}$C, $^{14}$N, $^{16}$O, $^{17}$O, $^{18}$O, $^{22}$Ne and $^{23}$Na. From Fig. \ref{fig:yield_comparison}, V20 produces higher $^{12}$C yields for the $2-3$\(M_\odot\) range, this is due to the earlier activation of TDU at 1.25\(M_\odot\) in V20, as opposed to the 3.5\(M_\odot\) minimum in this work. 

For $M>3$\(M_\odot\), the V20 models exhibit significantly higher temperatures at the base of the convective envelope. HBB initiates at a lower initial mass of 3.5\(M_\odot\) (rather than the 5.5\(M_\odot\) required in this work). A strong variance develops in yields of isotopes involved in hydrogen burning, such as $^{14}$N in Fig. \ref{fig:yield_comparison}. As discussed in V20, this can largely be attributed to the different treatment of convection between the two codes. V20 uses the full spectrum of turbulence model (FST) as opposed to the mixing length theory of convection used in the Monash code. The FST results in hotter temperatures at the base of the convective envelope, allowing for more efficient nucleosynthesis. Occurring alongside the onset of HBB, TDU efficiency drops from $\lambda$=0.48 (at 3.5\(M_\odot\)) to $\lambda$=0.15 at 4\(M_\odot\). This is reflected in the lower $^{12}$C yields for V20 at $M>3.5$\(M_\odot\) in Fig. \ref{fig:yield_comparison}. From \citet{Ventura05first} and \citet{Ventura12dust}, the authors discern the depressed $\lambda$ values are a combination of greater luminosities caused by stronger HBB, together with the heavily luminosity-dependent choice of mass loss approximation of \citet{Blocker95a}. 

\subsection{Stellar population yields}

The stellar yields calculated with Eq. \ref{Yield} provide us insight into the contributions of a single star to the interstellar medium. To decipher the chemical evolution of a galaxy, it is more useful to study the chemical contribution from a population of low and intermediate mass stars. We do this in a manner similar to \citet{Tinsley80} and \citet{Marigo01} by calculating yields of a stellar population, given by:

\begin{equation}
M^{\rm{pop}}_i  =  \frac{\int_{M_{\rm{low}}}^{M_{\rm{up (AGB)}}} f(m) \times m^{-2.35} dm}{\int_{M_{\rm{low}}}^{M_{\rm{up (tot)}}} m^{-2.35} dm }. 
\label{genEq}
\end{equation}

Here, $M_{\rm{low}}$ is the lower mass bound for a simple population, taken in this work to be $M_{\rm{low}}=1$\(M_\odot\). $M_{\rm{up (AGB)}}$ is the upper mass bound for AGB stars and $M_{\rm{up (tot)}}$ the total upper mass bound for the population, given by values of 8 and 100\(M_\odot\), respectively. In the numerator of Eq. \ref{genEq}, $f(m)$ is a function we generate by interpolating the data points of the individual yields. This represents how the stellar yields vary as a function of mass for a given metallicity. $f(m)$ is weighted with the initial mass function (IMF) \citep{Salpeter55}. The stellar IMF describes the number density of stars on the zero-age main sequence in a population as a function of stellar mass. \citet{Salpeter55} measured the IMF to be a power law for masses between 0.4 and 10\(M_\odot\), $\xi(m)=Cm^{-x}$, where $x$=2.35. On the denominator, we simply integrate the IMF again, however now for the total mass within the population. $M^{\rm{pop}}_i$ then represents the chemical contribution (of species $i$) produced by low and intermediate mass AGB stars, proportional to the total mass distribution in a population. 

Other IMFs provide slightly different power law slopes compared to Salpeter, along with a turnover for the lowest mass stars (e.g. \citet{Kroupa93}). The turnover refers to the over prediction of low mass stars ($M<0.5$\(M_\odot\)) but this is outside of the mass range of our models. \citet{Kroupa01} measured a similar power law for stars with $M>0.5$\(M_\odot\), $x=2.3\pm 0.7$ with a 99\% confidence interval. Power laws within this range can reproduce the mass distribution in Milky Way type spiral galaxies \citep{Zhang18}. The uncertainty in measuring stellar IMFs increases when changing the focus galaxy from local spiral disks to distant starburst galaxies. These rapid star forming galaxies require a top heavy stellar IMF to reproduce the metallicity distributions within the galaxy \citep{Ballero07, Zhang18, Marks12} which suggests that there are more supernovae events than a Salpeter power law predicts. This is more likely to impact a higher mass range than the yields we cover in this paper (i.e. stars massive enough to undergo core collapse supernovae). To pursue this further is beyond the scope of our research, but we emphasize this is a notable uncertainty and could the increase the importance of the most massive of our mass range, super-AGB stars. 

\begin{figure*}
	\includegraphics[width=16cm]{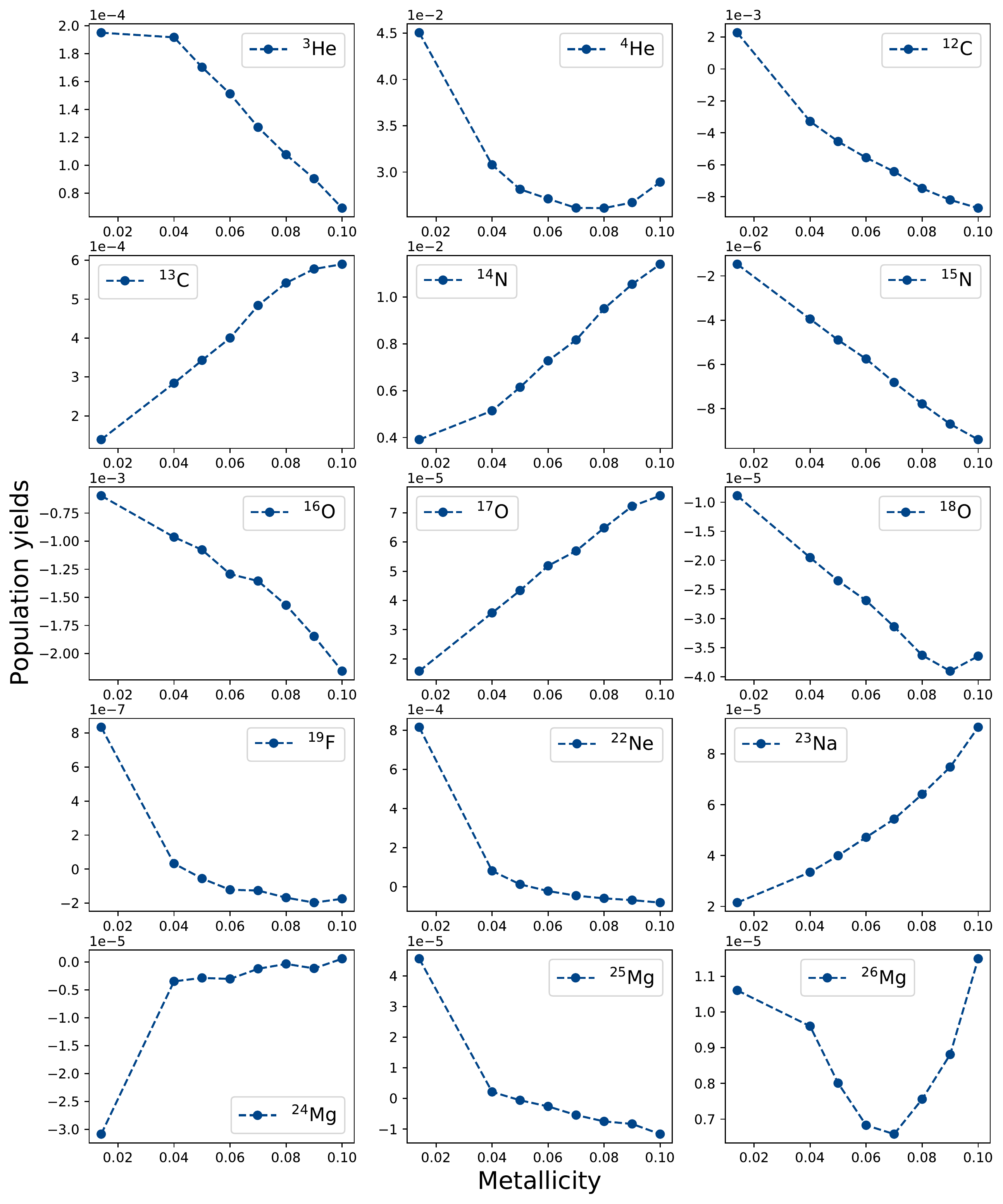}
    \caption{Population yields. Many of the same trends are observed in these plots as we found in sub section \ref{stellaryields}, weighted to a lower mass regime.}
    \label{fig:stellargen1}
\end{figure*}

Figs. \ref{fig:stellargen1} and \ref{fig:stellargen2} show the population yields from our calculations, they cover the same range of isotopes discussed in \S~\ref{stellaryields}. These plots reflect many of the same trends previously discussed, with the major difference being that they are weighted towards the lower mass range. For example, the population yields of $^{3}$He in Fig. \ref{fig:stellargen1} show a decreasing trend with increasing metallicity. This correlates to the individual $^{3}$He yields in Fig. \ref{fig:nonweighted_1} for the low mass range. However, for intermediate mass models with $M>4.5$\(M_\odot\), the $Z=0.014$ models produce the lowest quantity of $^{3}$He in Fig. \ref{fig:nonweighted_1}. Due to the excess of low mass stars in a stellar population, Fig. \ref{fig:stellargen1} indicates that $Z=0.014$ has the highest production of $^{3}$He, irrespective of its behaviour at 4.5\(M_\odot\) and greater.

Figs. \ref{fig:stellargen1} and \ref{fig:stellargen2} show that there are clear decreases with increasing metallicity in the isotopes $^{3}$He, $^{4}$He, $^{12}$C, $^{15}$N, $^{16}$O, $^{18}$O, $^{25}$Mg, $^{27}$Al, $^{22}$Ne and $^{19}$F. We predict that the yields of isotopes produced in the burning shells and the intershell region will decrease with increasing metallicity, given that they rely on efficient TDU to mix the products up to the surface. With the conservation of CNO elements, higher levels of CNO ingredients in the initial gas of high metallicity stars results in higher destruction of $^{15}$N, $^{16}$O and $^{18}$O in metal-rich stars. This is consistent amongst the entire mass range and so transfers to the population yields. 

Consistent increases with increasing metallicity are observed in the isotopes $^{13}$C, $^{14}$N, $^{24}$Mg, $^{17}$O and $^{23}$Na. For the same reason that we have a moderate destruction of $^{15}$N, $^{16}$O and $^{18}$O with increasing Z, we have increasing production of $^{14}$N, $^{13}$C and $^{17}$O. There is a large net increase in $^{23}$Na as metallicity increases. The strong metallicity dependence that favours an increasing metallicity is observed in both AGB stars and type II supernovae, the other prevalent site of sodium production \citep{Vaughan18, Kobayashi06, Kobayashi20, Ventura13}. $^{24}$Mg is destroyed at lower metallicities, this tapers off for high metallicity models that do not reach temperatures sufficient to deplete $^{24}$Mg. 

\subsection{Production of heavier isotopes in high metallicity models}
\label{yieldsResults_sprocess}

\begin{figure*}
	\includegraphics[width=16cm]{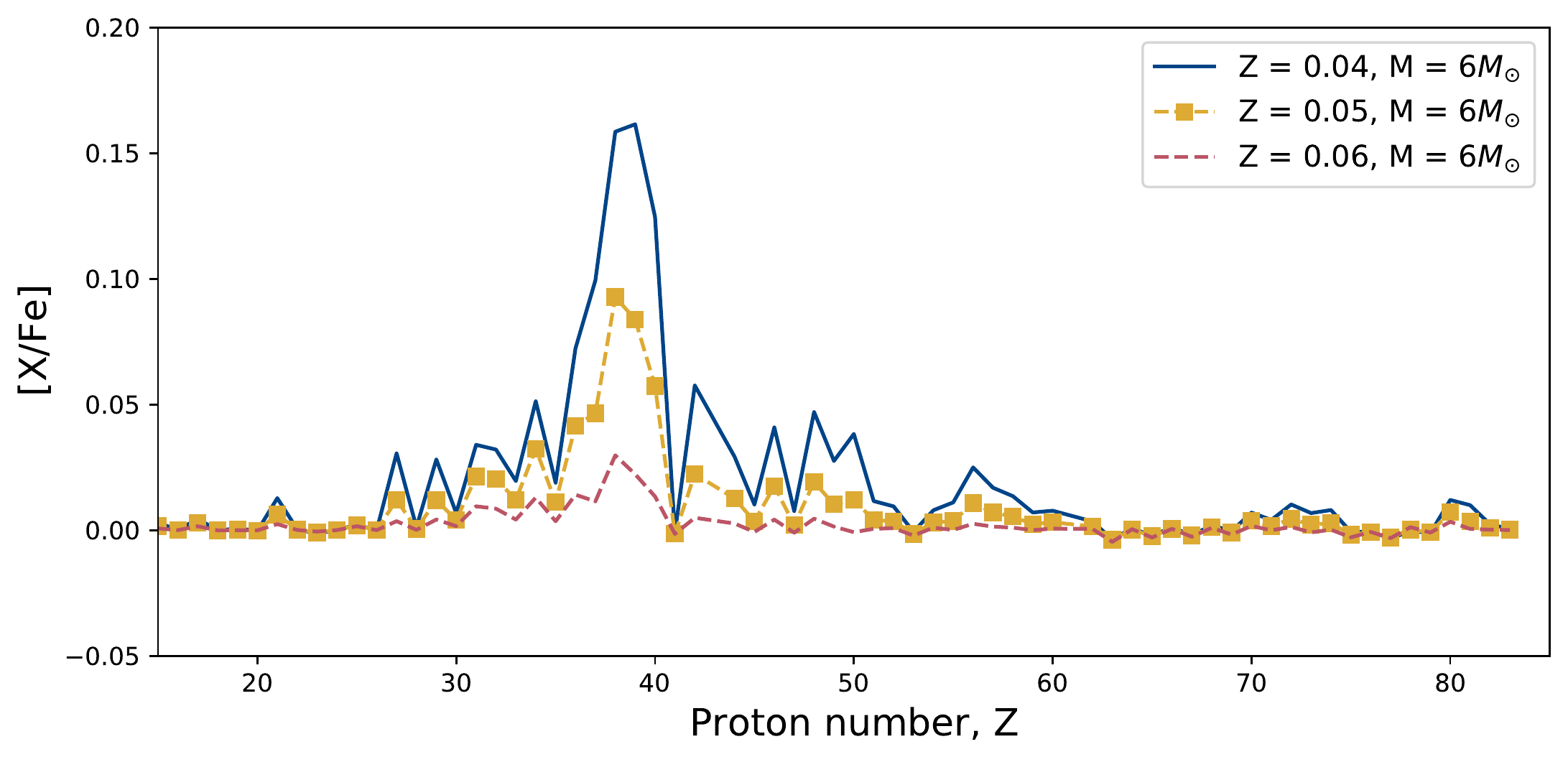}
    \caption{[X/Fe] ratios of the stellar yields for 6\(M_\odot\) models with $0.04 \leq Z \leq 0.6.$}
    \label{fig:sprocess}
\end{figure*}

Fig. \ref{fig:sprocess} shows the stellar yield [X/Fe] ratios for elements heavier than iron predicted by our 6\(M_\odot\) models of $Z=0.04, 0.05$ and $0.06$. These abundances depend not only on the efficiency of the \textit{s}-process in high metallicity models, but also on the efficiency of TDU and the level of dilution that will occur with the large envelopes of intermediate mass stars. 

The \textit{s}-process produces peaks in abundance that follow the magic nuclei $^{88}$Sr, $^{138}$Ba and $^{208}$Pb. As outlined in \citet{Lugaro16}, these nuclei are particularly stable to neutron capture and therefore act as bottlenecks. The level of production at each peak (and the ratios between the peaks) give us insight into the neutron exposure within the star. The neutron exposure defines the integral of the neutron flux with respect to time. We find most heavy element production occurs around the first \textit{s}-process peak at strontium (Z=38). The size of the peak decreases as metallicity reaches $Z=0.06$. Production around the first peak is expected for solar and super solar metallicities \citep{Karakas16}. The neutron exposure is proportional to the number of neutrons per iron seed. Given that higher metallicity gases, by definition, contain more iron seeds, more neutrons are required to produce the same abundance peak found in a lower metallicity model. Thus, even with the addition of partial mixing zones to assist the activation of the $^{13}$C neutron source, the neutron exposure is so low that abundances only reach the first \textit{s}-process peak. 

Each of the 6\(M_\odot\) models reach maximum temperatures in the intershell region that surpass 300 MK. This suggests that the $^{22}$Ne neutron source also makes contributions to our models. Although activated at 300 MK, the $^{22}$Ne neutron source operates most efficiently for T $>$ 300 MK \citep{Fishlock14}. The highest temperature reached is 330 MK, approached for the last $\approx$ 5 TPs of the $Z=0.04$ model. At $Z=0.05$, the maximum temperature reached declines to approximately 320 MK for $\approx$ 3 TPs, further to 310 MK for $Z=0.06$ for only 1 TP. As a result, there is likely some contribution of the $^{22}$Ne neutron source for all three models, but this becomes increasingly mild as metallicity increases.

\section{Discussion}
\label{Discussion}

Our results show that metal-rich stars are characterized by very low levels of helium burning products and \textit{s}-process nucleosynthesis with relatively high quantities of hydrogen burning products. This follows from the extremely inefficient mixing processes on the TP-AGB as well as mixing processes that occur prior to the TP-AGB. In this section, we will attempt to provide further insight as to why variances in the nucleosynthesis of AGB stars occur as metallicities increase beyond solar. 

\subsection{Third dredge up} 
\label{TDU}

Very low quantities of He-shell burning products are mixed to the surface of our AGB models. This indicates that TDU becomes much less efficient as metallicity increases. Quantitatively, we can measure this change using the TDU efficiency parameter, $\lambda$. $\lambda$ quantifies the ratio of the mass brought to the stellar surface by the convective envelope, $\Delta M_{\rm{dredge}}$, to the amount with which hydrogen burning increases the core mass during the previous interpulse period, $\Delta M_{c}$ \citep{Karakas02}:

\begin{equation}
\lambda = \frac{\Delta M_{\rm{dredge}}}{\Delta M_{c}}. 
\end{equation}

\begin{figure*} 
	\includegraphics[width=16cm]{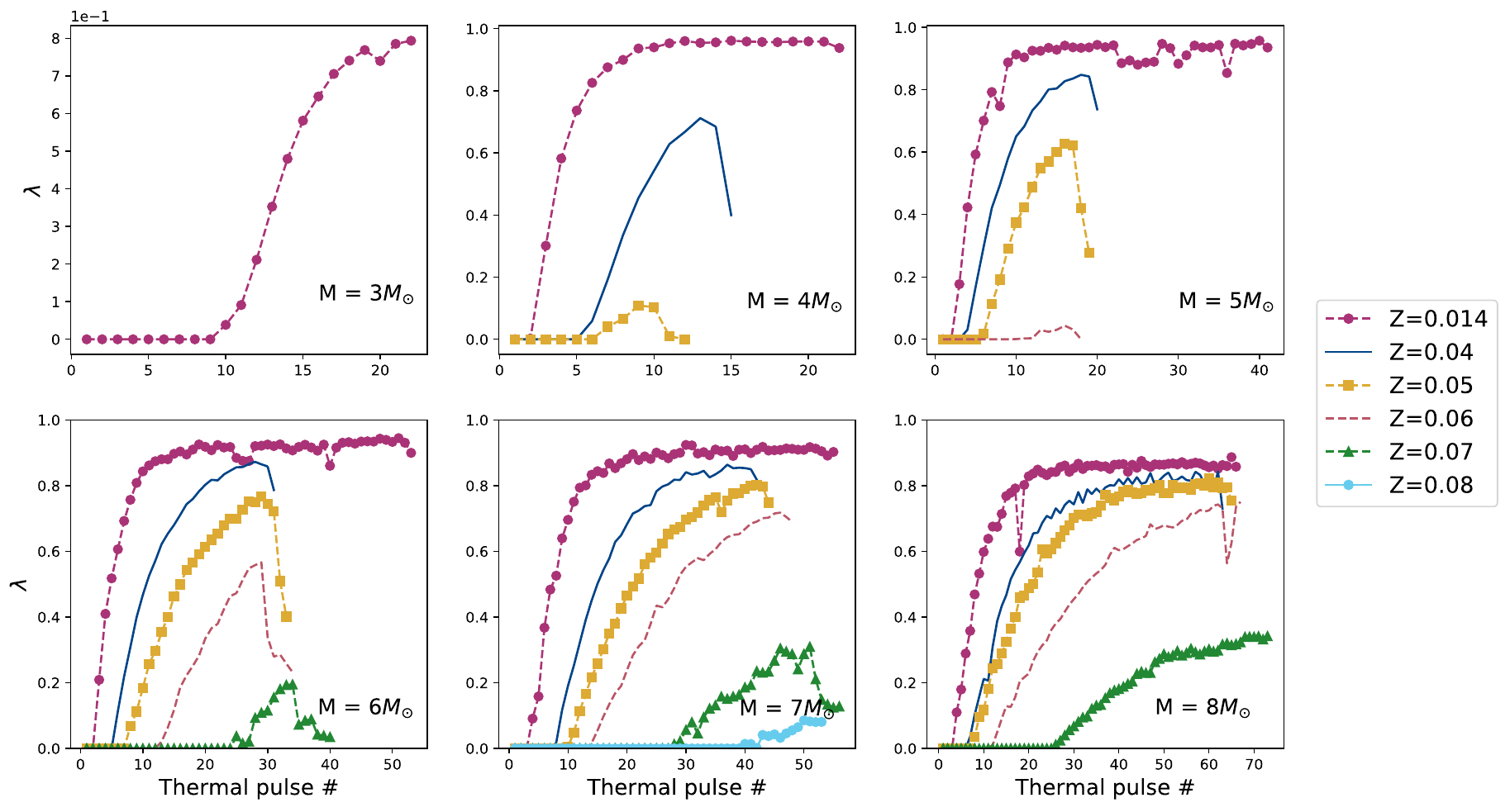}
    \caption{TDU efficiency decreases with increasing metallicity. This is demonstrated by following the TDU efficiency parameter, $\lambda$, which is defined in Sub Section \ref{TDU}. The mass threshold for TDU increases with increasing metallicity, at 3\(M_\odot\) TDU is significantly active only for $Z=0.014$. The intermediate mass range boasts decent values of $\lambda$ for the entire metallicity range.}
    \label{fig:lambda_max}
\end{figure*}

There is a strong metallicity dependence for TDU to operate, with a decreasing metallicity favouring higher TDU efficiency \citep{Boothroyd88, Karakas02, Straniero03}. This is reflected in our models \citep{Karakas21}. In Fig. \ref{fig:lambda_max} we show $\lambda$ as a function of TP number for $3-8$\(M_\odot\) models with varying metallicities. This depicts the stark contrast between the TDU efficiency of solar versus super solar metallicity stars. The models with solar metallicity reach $\lambda\approx$1 early, spending most of their time of the TP-AGB at high efficiencies. The $Z=0.04$ and $0.05$ models of equal mass struggle to reach anywhere near the same $\lambda$ value. Due to their increasingly short lives on the AGB, once they have reached their respective maximum, $\lambda$ begins to almost immediately decline again. 

\begin{figure} 
	\includegraphics[width=\columnwidth]{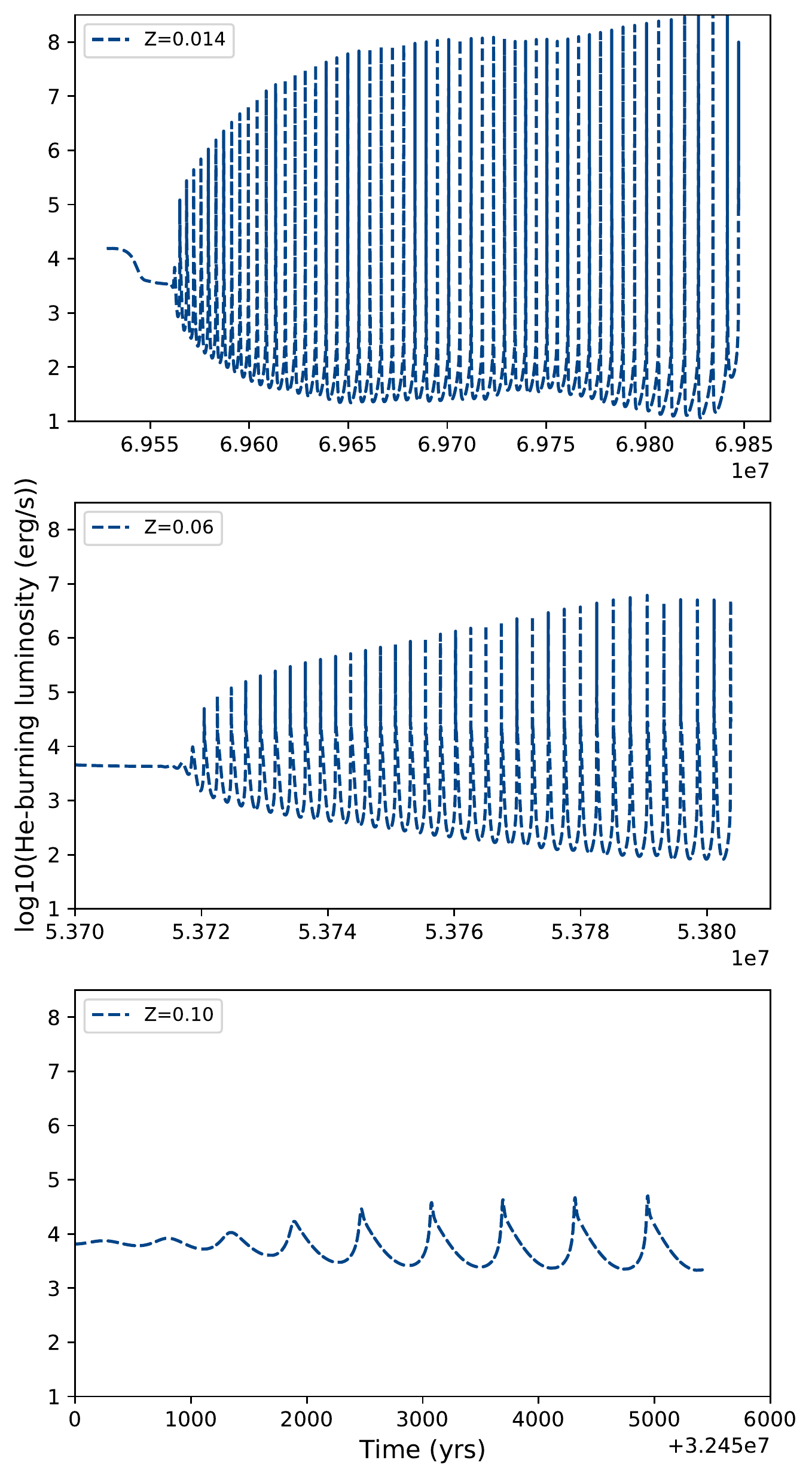}
    \caption{Helium burning luminosity for 6\(M_\odot\) of $Z=0.014, 0.06$ and $0.10$.}
    \label{fig:LHe_all}
\end{figure}

The efficiency of TDU is linked to the strength of helium shell flashes and how variables such as opacity, core mass, initial stellar mass and the temperature at the base of the convective envelope during a thermal pulse change as a consequence of the variances in initial composition \citep{Karakas02, Marigo99, Wood81}. The helium burning luminosity is the most likely cause given that it consistently decreases with metallicity. Fig. \ref{fig:LHe_all} shows the decreasing magnitude of the helium burning luminosity for models of 6\(M_\odot\) for $Z=0.014, 0.06$ and $0.10$. We observe an indirect relationship between $\lambda$ and the magnitude of the helium burning luminosity, given that the size of the thermonuclear runaway is proportional to the extent with which the outer layers of the star expand \citep{Pols01thermal, Frost96}. This suggests that for our higher metallicity models, there is less thermal energy available to convert to mechanical energy, which results in a shallower dredge up. 

There are a few disadvantages to our highest TDU efficiency values occurring in intermediate mass stars. Firstly, the intershell regions of low mass stars are more massive than those of intermediate mass by roughly an order of magnitude. This in essence means that regardless of $\lambda$ efficiency, less material will be dredged into the envelope for an intermediate mass star. The other shortcoming is the dilution of that smaller quantity of material into a massive envelope. This is particularly relevant in analysing our \textit{s}-process nucleosynthesis results as shown in Fig. \ref{fig:sprocess}, given that the observation of heavy elements (and production, for the case of the $^{13}$C neutron source) requires dredge up.

\begin{figure}
	\includegraphics[width=0.9\columnwidth]{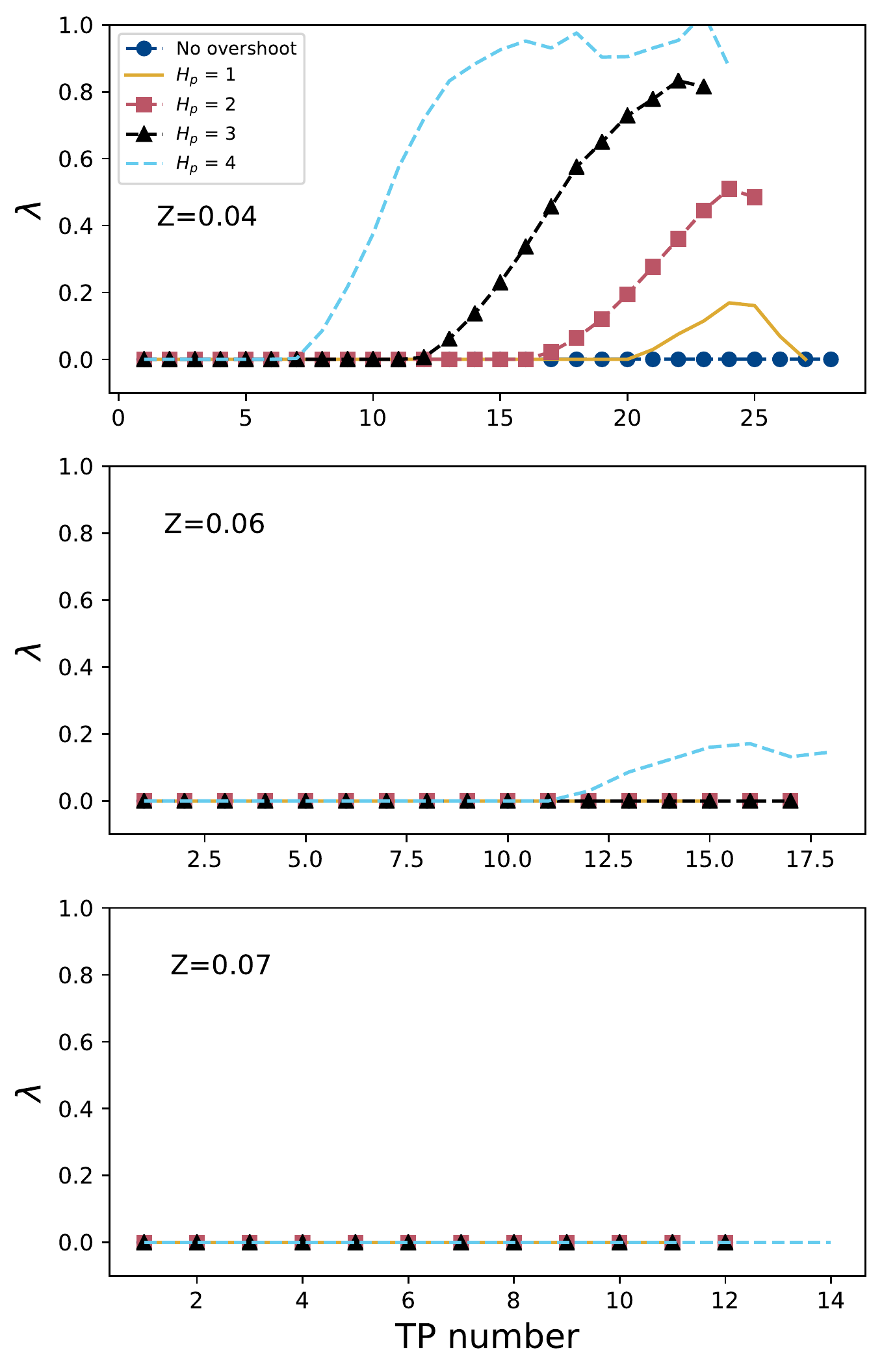}
    \caption{Investigating the effects of convective overshoot on 3\(M_\odot\) models. Results for the $Z=0.04$ models are displayed in the top panel, $Z=0.06$ in the centre panel and $Z=0.07$ in the lower panel. The navy-blue tracks with circles indicate our original 3\(M_\odot\) models with no overshoot. We emphasize here that normally (i.e. with no forced overshoot) there are virtually no TDU events in this mass range for $Z\geq0.04$. When we include overshoot, we can generate quite significant mixing efficiency for $Z=0.04$ models by increasing the pressure scale height, $H_p$, from 1–4. At higher metallicities of $Z=0.06$, we require $H_p$=4 to force very inefficient mixing. $Z=0.07$ acts as our metallicity ceiling for the low mass range, where no mixing can occur, even with $H_p$=4. These conclusions are still based on assumptions (primarily, that $H_p$=4 is a suitable and realistic constraint). However, it allows us to quantitatively conclude that there is likely an uncertainty of $Z=\pm0.01$ in deciding the metallicity ceiling for TDU for a given mass range.}
    \label{fig:overshoot_comp}
\end{figure}

Determining the boundaries of convective regions in our models directly impacts the occurrence and efficiency of TDU in our calculations. If we underestimate the extent to which the convective envelope can reach into the central depths of the star, we will significantly reduce the quantity of helium burning and \textit{s}-process nucleosynthesis products that we will calculate for the stellar surface. As we discuss in \citet{Karakas21}, we do not include formal overshoot in our models, such as convective boundary mixing \citep[e.g.][]{Herwig00, Davis19, Cristallo15}. We have run a few exploratory calculations of 3\(M_\odot\), $Z=0.04-0.07$ with convective overshoot as described in \citet{Karakas10mixing}. In Fig. \ref{fig:overshoot_comp}, we show the TDU efficiency as a function of TP number for varying metallicities and pressure scale heights ($H_p$) of $H_p$=1 to 4. Without overshoot, we find no TDU at any metallicity for the low mass range. With overshoot values of $H_p$=1 to 4, we find that we can generate TDU episodes for 3\(M_\odot\) at metallicities of $Z=0.04-0.06$, with a metallicity ceiling of $Z=0.07$. Thus, if we were to include overshoot with similar pressure scale height values to \citet{Kamath12}, it is very likely that the convective envelope is capable of reaching further into the central regions of our models than we have shown. This would increase both the number and efficiency of TDU events, as well as extending the maximum metallicity ceiling for which TDU events can occur. We emphasize that our overall conclusions would not change, the quantity of helium burning and \textit{s}-process nucleosynthesis products found on the stellar surface will still decrease with increasing metallicity. However, the absolute levels of these isotopes may be higher than what we have calculated.

\subsection{Proton capture nucleosynthesis}

In contrast to the low levels of helium burning products, our yields show very high levels of proton capture nucleosynthesis. These emanate from various sources; HBB on the TP-AGB, SDU on the early-AGB and FDU on the first giant branch. 

\begin{figure*}
   \centering
   \includegraphics[width=14cm]{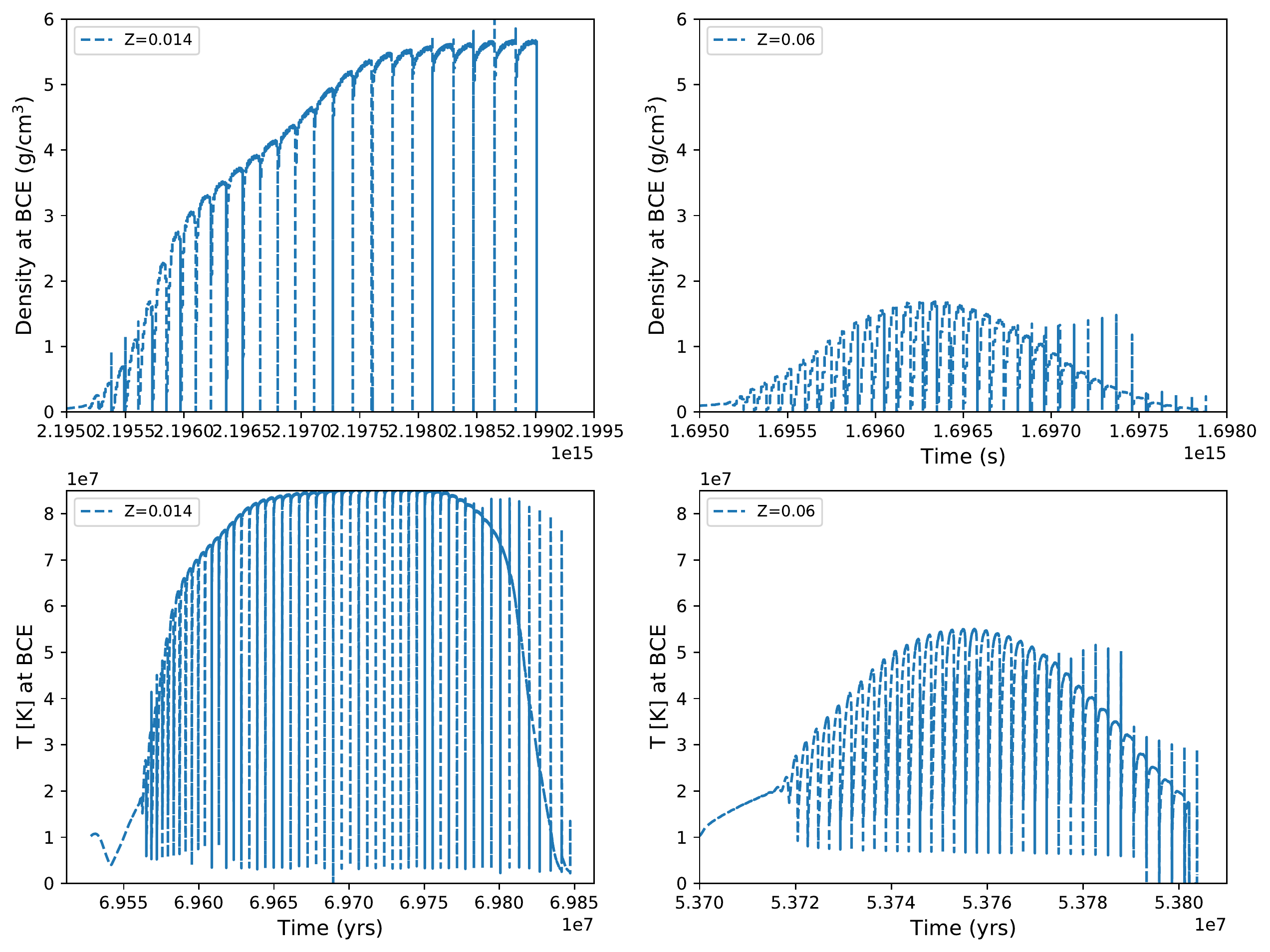}  
   \caption{Top panel: density at the border of the convective region. Given the higher opacities of very metal-rich stars, they are much less compact than their metal poor counterparts. This is emphasized by the significantly lower densities in this region for the $Z=0.06$ model in comparison to the solar metallicity model. Bottom panel: Temperatures at the base of the convective envelope. As a consequence of the much lower density in this region at $Z=0.06$, the associated temperature peaks are equally lower than for $Z=0.014$.}
   \label{fig:dens_tbce}
\end{figure*}

\begin{figure}
   \centering 
   \includegraphics[width=\columnwidth]{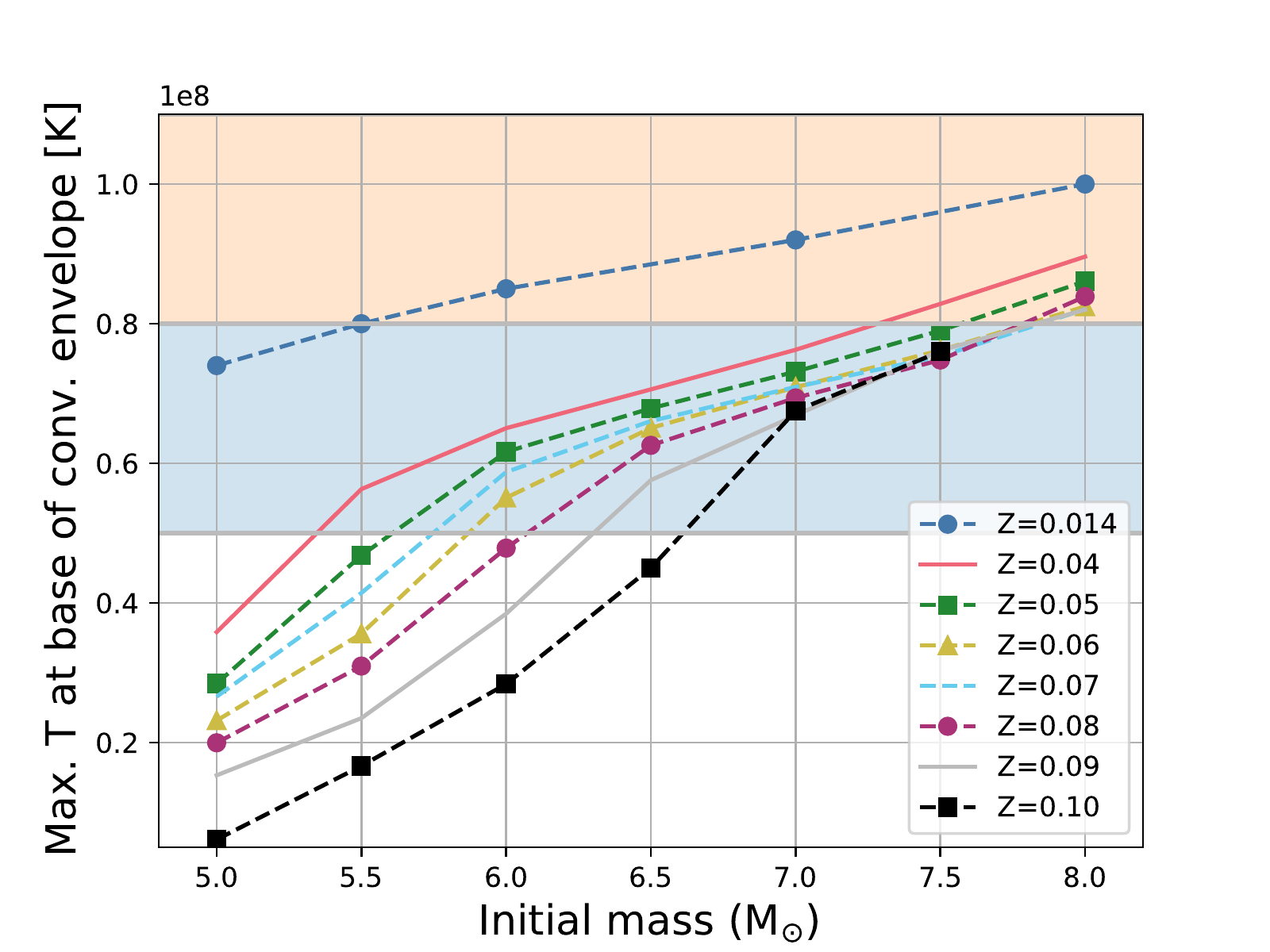}  
   \caption{The maximum temperature reached at the base of the convective envelope as a function of initial mass. At $\approx$ 0.5 $ \times 10^8$K, temperatures are sufficiently high to begin CN burning at the base of the convective envelope (blue shaded region). Above $\approx$ 0.8 $ \times 10^8$K, temperatures may be high enough for full CNO burning and possibly the Mg-Al and Ne-Na chains to operate (orange shaded region).}
   \label{TBCEmax}
\end{figure}

Whether burning will or will not occur at the base of the convective envelope is dependent on the temperatures reached at this location. Metal-rich stars are much puffier than their compact, lower metallicity counterparts. This is due to the positive relationship between the opacity of stellar gas and initial metal content. In Fig. \ref{fig:dens_tbce}, we show the density and temperature at the base of the convective envelope for the 6\(M_\odot\) models of solar metallicity and $Z=0.06$. Since the metal-rich model is more extended, the density at the base of the convective envelope significantly decreases with increasing metallicity, where the solar metallicity model has a peak value nearly 3.5 times the value of the $Z=0.06$ model. As a result, Fig. \ref{fig:dens_tbce} shows the significantly lower temperatures in the $Z=0.06$ model. The decreasing temperature at the base of the convective envelope with increasing metallicity is consistent across the entire mass range. Fig. \ref{TBCEmax} exhibits the maximum temperature reached at the base of the convective envelope as a function of mass for each metallicity. At $T\approx 0.5\times 10^8$K, temperatures are sufficiently high to begin CN burning at the base of the convective envelope, represented by the blue shaded region. Above $T\approx 0.8 \times 10^8$K, temperatures are high enough for full CNO burning, represented by the orange shaded region. As $T$ increases beyond $0.8-1.0\times 10^8$K the Mg-Al and Ne-Na chains may operate \citep{Forestini97, Ventura13, Karakas03}. As metallicity increases, the peak temperature values decrease. 

From Fig. \ref{fig:lambda_max}, we see that as Z increases, the maximum $\lambda$ values reached also decrease. This is key to distinguish between primary and secondary production from hydrogen burning. Low metallicity stars rely on TDU to bring fresh, primary carbon into the envelope to produce primary nitrogen. High metallicity stars experience much less efficient TDU and thus to an extent rely on their inherent higher levels of carbon to make secondary nitrogen in the envelope. This is important given that primary nucleosynthesis will increase the global metallicity of the stellar gas expelled from a star over its lifetime, contributing fresh metals back to the host galaxy over time. Secondary nucleosynthesis is simply re-arranging the initial isotopic composition of the gas, providing no fresh metals to the galaxy.

For the highest metallicity models, $Z=0.10$, the temperatures at the base of the convective envelope are only high enough for mild HBB to occur in the 7 and 7.5\(M_\odot\) models. Yet we can see in Fig. \ref{fig:nonweighted_1} that the $Z=0.10$ models produce the largest quantities of $^{14}$N over the span of a lifetime. In this case, proton capture levels will be exclusively due to FDU on the red giant branch for low mass stars. FDU mixes the products of partial hydrogen burning from the main sequence to the surface. For intermediate mass stars, this will be a combination of FDU and SDU, where SDU mixes the products of full hydrogen burning to the surface on the early-AGB. From \citet{Karakas21} we found that mixing episodes prior to the thermally pulsing AGB, SDU in particular, are relatively independent of metallicity. Consequently, the high proton-capture yields of the highest metallicity models likely represent secondary production. For the models that undergo TPs and some TDU, $Z=0.04-0.08$, their yields will be some combination of both primary and secondary production in their envelopes. 

\subsection{Presolar grains}

In this work, we have solely addressed the composition of \textit{gas} expelled from the most metal-rich stars. Also important to their story is the composition and behaviour of \textit{dust and grain} production in the circumstellar environment. The main sources of pre-solar grains are AGB stars and core collapse supernovae \citep{Zinner14grains, Nittler16grains, Gallino90SiC}. By studying the isotopic composition of these grains, we are provided with another way to constrain our stellar evolution and nucleosynthesis modelling. None of the models in this work become carbon rich, so they are likely to produce dust comprised mainly of oxygen rich silicates. However, given the uncertainties in modelling third dredge up (see section \ref{TDU}), it is possible that stars with $Z \geq 0.04$ could expel carbon rich winds, forming grains such as silicon carbides (SiC). Although there are various types of pre-solar grains, SiC grains are the most well studied in the literature. The majority of large SiC grains were formed in carbon rich stars of super solar metallicities \citep{Lugaro18meteoritic, Lugaro20origin}. \citet{Lugaro18meteoritic} conclude that these higher than solar metallicity models (in their work, $Z_{\rm max}=0.03$) address the Zr, Sr and Ba isotopic ratios in the grains very well.

The parent stars of SiC grains require stellar surfaces with C/O $\geq 1$, which is a necessary condition for the grains to condense. The Monash models of $Z=0.03$ referenced in \citet{Lugaro20origin} (originally published in \citet{Karakas14He}) were able to become carbon rich for masses between $2.5-4$\(M_\odot\) when including a small amount of convective overshoot (as described in \citet{Karakas16}). Whether or not AGB models are able to satisfy the surface C/O $\geq 1$ condition depends heavily on the treatment of TDU and mass loss rates. The models in this work with $Z \geq 0.04$ experience less efficient TDU at higher masses than their solar and $Z=0.03$ metallicity counterparts from \citet{Karakas14He} and endure much shorter lifetimes in general. Consequently, none of the $Z=0.04$ models here can reach C/O $\geq 1$ at the surface. As discussed in section \ref{TDU}, it is possible that we have measured a lower limit for TDU in our models given that we do not include convective overshoot. Combined with the uncertainties in mass loss rates, this may mean that some of our $Z=0.04$ models could become carbon rich if they experienced more efficient TDU and longer lifetimes on the TP-AGB. To explore this further is beyond the scope of the paper, however we do note that both \citet{Nanni14evolution} and \citet{Ventura20} studied dust production at super solar metallicities ($Z=0.04-0.06$ for the former, $Z=0.03-0.04$ for the latter). The $Z=0.03$ models in \citet{Ventura20} reached C/O $\geq 1$ for the mass range $2-3.5$\(M_\odot\), the 2.5\(M_\odot\) $Z=0.04$ model also achieves this. In \citet{Nanni14evolution}, only the 5\(M_\odot\), $Z=0.04$  model reaches C/O $\geq 1$. Both found that the majority of models with metallicities of $Z \geq 0.04$ completed their evolution as oxygen rich, with dust dominated by silicates.

\subsection{Uncertainties and limitations of our work}

The primary uncertainties that impact stellar nucleosynthesis calculations are mirrored in those of stellar evolution calculations. These are discussed in \citet{Karakas21} and cover mainly the mass loss rates for the giant branches, numerical methods for treating convection and determination of convective boundaries. Additional uncertainties accompany the large reaction networks we use in nucleosynthesis codes which can have a significant impact on the final surface abundances \citep[e.g.][]{Izzard07, Van08, Lugaro04reaction, Herwig06nuclear}. In this section we will discuss how these uncertainties can directly affect our yields.  

\subsubsection{Mass-loss rates}

In our models we use the mass loss approximation of \citet{Reimers75} for the red giant branch and \citet{Vassiliadis93} for the AGB. \citet{Vassiliadis93} was calibrated from AGB stars in our Galaxy (approximately solar metallicity) and in the Magellanic Clouds, which have metallicities a factor of 3-5 less than solar. It is unclear how applicable these approximations are to very metal-rich stars. Higher mass loss rates will significantly reduce the number of TPs, while a lower mass loss rate will extend the AGB lifetime. The consequence of a lower mass loss rate on the stellar yields could mean more TDU episodes, this would result in higher levels of chemical enrichment. For our most metal-rich models, an increase in the number of TPs does not likely lead to more-or indeed any-TDU episodes. Regarding HBB, the faster the outer envelope is eroded, the faster the temperatures at the base of the envelope will cool, which will inhibit any burning at the base of the envelope. In \citet{Karakas21}, we looked at the example of the $Z=0.10$ case. With the standard mass loss rates used, only 3 of the models at this metallicity (5, 5.5 and 6\(M_\odot\)) reach the thermally pulsing stage and only the 7 and 7.5\(M_\odot\) models reach temperatures sufficient for HBB on the early-AGB. Without mass loss, the 6.5\(M_\odot\) model achieves 37 thermal pulses and reaches maximum temperatures capable of HBB (75 MK). This example is an extreme case, but it does show that lower mass-loss rates on the AGB can lead to more TPs and in the case of the 6.5\(M_\odot\), the activation of HBB \citep[e.g.][]{Ventura05}.

\subsubsection{Lack of calibration for modelling approximations in metal-rich stars}

A further uncertainty, relatively unique to our work in metal-rich nucleosynthesis, is that most of the theoretical assumptions we use are calibrated for solar or metal-poor observations. Given that the primary goal of theoretical research is to physically represent what is seen in nature, it is a crucial step that our models are constrained to observation. However, the difficulty of observing AGB stars is almost on par with the challenges in modelling these stars. This is both due to low numbers of AGB stars (given the short timescale of the AGB) and difficulty in resolving individual stars in the infrared \citep{Boyer19}. There are some metal-rich AGB stars that have been resolved in the bulge of the Milky Way, for example 2M17445261-2914110 and 2M17452187-2913443 \citep{Schultheis03, Schultheis20} which have respective metallicities of [M/H]=0.13 and [M/H]=0.21. 
An effective method to constrain theoretical AGB models that avoids resolving actual AGB stars is to compare theoretical abundances to those detected in planetary nebulae (PNe). A PNe is comprised of a dense stellar remnant surrounded by ionized gas, of which is made up of the outer layers of the progenitor star. Thus, the composition of the ejected gas is a direct reflection of the nucleosynthetic and mixing processes the star endured, primarily on the AGB. Type I PNe show high He/H and N/O ratios that are signatures of proton-capture nucleosynthesis \citep{Karakas09}. Consequently, Type I PNe provide a good constraint on the efficiency and extent of HBB (and the lack of TDU). An example of these studies is \citet{Ventura15}, who compared theoretical AGB models to the abundances of PNe in the Large Magellanic Cloud (LMC). An example of their findings was that their $M\geq6$\(M_\odot\) models, with surface abundances dominated by HBB, quite accurately reproduced the abundances of the nitrogen rich PNe in their sample. There is extensive observational data available for the LMC, however given the average metallicity of the stellar populations within are sub-solar \citep{Carrera08} this is not comparative or useful for this work. PNe studies of metal-rich regions centre around the Galactic Bulge, which has a higher average metallicity than the disk of the Milky Way. For example, \citet{Perea09} studies the dual-dust chemistry phenomenon using data from the Galactic Bulge, investigating the link between carbon rich PNe to possible low mass, oxygen rich AGB progenitors. Heavy element abundances in PNe also provide useful constraints on the \textit{s}-process. For example, \citet{Lugaro16stellar} suggest Krypton and Xenon observations could be used as a proxy for the [hs/ls] \textit{s}-process index. 

\subsubsection{Extra mixing on the red and asymptotic giant branches}

It is important to note that the low mass yields of some light elements, particularly $^{3}$He and $^{4}$He (including a few others such as $^{7}$Li that we do not discuss), are likely affected by extra mixing mechanisms prior to the TP-AGB that we do not include in our calculations. Standard nucleosynthesis calculations (such as that used in this paper) can significantly over-estimate the quantity of $^{3}$He produced in comparison to observations of these isotopes in the Galaxy \citep{Rood76, Lagarde12}. It has since been proposed that extra mixing is necessary to reduce the amount of $^{3}$He produced in low mass stars, this is likely linked to thermohaline mixing on the red giant branch \citep{Rood84, Ulrich1972, Charbonnel07, Angelou11thermohaline}. Galactic chemical evolution models with input yields that include thermohaline and rotational mixing match helium observations more closely than those without \citep{Lagarde12, Lagarde11thermohaline2}. However, it is important to note that the efficiency of both mixing mechanisms (rotation induced and thermohaline) decrease with increasing metallicity (for further details see \citet{Charbonnel10thermohaline1, Lagarde11thermohaline2}). Also, the magnitude of production (and destruction) of some light isotopes is highly dependent on the numerical details of how each code models thermohaline mixing and the choice of diffusion coefficient values \citep{Lattanzio15, Angelou15, Henkel2017}. Consequently, it is not clear how much of an impact this extra mixing has on our very metal-rich, low mass models.

Whilst there is general agreement amongst the literature that some extra mixing is required on the first giant branch, the consensus is mixed as to whether this also applies to the AGB. Unlike the red giant branch, extra mixing on the AGB is unlikely to be a result of thermohaline mixing. \citet{Stancliffe10} shows that although thermohaline mixing can produce lithium-rich carbon stars, it does not operate for long enough on the TP-AGB to significantly lower the $^{12}$C/$^{13}$C ratio. \citet{Karakas10mixing} can explain the $^{12}$C/$^{13}$C ratios of most AGB stars using models that assume extra mixing on the red giant branch alone. Some AGB stars have very low $^{12}$C/$^{13}$C ratios that are inconsistent with their results. Their models remain lithium poor, however they note that roughly 10\% of AGB stars are both carbon and lithium rich. This suggests extra mixing may operate in a small fraction of the population. \citet{Busso10deepmixing} require extra-mixing in their models to constrain oxygen and aluminium ratios in pre-solar grains. They suggest that the extra-mixing is driven by magnetic buoyancy, supported by \citet{Palmerini21grains}. More recently, \citet{Abia17mixing} find that although extra mixing on the AGB may be required for some stars, it is likely to be rare. What these research groups do agree on is that extra mixing is likely to be much more efficient at metallicities lower than solar. Consequently, again it is unclear of the impact extra mixing would have on our models.

\subsubsection{Further uncertainties in $^{13}$C pocket formation}

In Section \ref{methods_heavyelements}, we briefly discussed the various methods available in the literature to form a $^{13}$C pocket. In this work we insert an artificial partial mixing zone. We refer the reader to \citet{Lugaro18meteoritic} for a very detailed discussion of the proton profile used in the Monash code, depicted in Fig. 1 of their paper. Briefly, we use an exponential profile to describe the extent of mixing from the envelope into the intershell region. This reaches down to a chosen mass extent (in this work that is $1 \times 10^{-4}$\(M_\odot\)). Abundances of $^{13}$C and $^{14}$N are then determined by $^{12}$C and $^{13}$C proton capture rates. There is some concern of the impact of the chosen proton profile on $^{13}$C and $^{14}$N abundances, and the downstream impact this can enact on the efficiency of the \textit{s}-process and some lighter elements such as $^{22}$Ne. \citet{Buntain17partialmixing} explored the impact of the proton profile in an artificially implemented partial mixing zone. For models with metallicities around solar (where $^{13}$C pockets are burned radiatively), they find that the impact of changing the mixing profile produces a very similar effect to that of changing the mass extent of the partial mixing zone (e.g. \citet{Fishlock14}). Both impact \textit{absolute} abundances, but the \textit{relative} distribution of the [Ba/Sr] and [Pb/Ba] ratios remains the same \citep{Buntain17partialmixing}. Consequently, similar to our discussions of TDU efficiency and yield trends in general-the absolute values of the \textit{s}-process elements may vary, but the relative ratio of the peaks should be similar.

\subsubsection{Single star nucleosynthesis}

Lastly, there are undeniably limitations to studying the nucleosynthesis of single stars, given that many stars (if not the majority) are expected to be part of a binary system. If our post-AGB stars are able to accrete enough mass from a companion star to reach the Chandrasekhar mass, the resulting type Ia supernovae will release these previously locked metals to pollute the surrounding interstellar medium \citep{Travaglio15, Maoz14}. Binary systems rely on the complex interplay between far more initial factors than are required for single star calculations \citep{Izzard06} and extend beyond the scope of this work, but this remains an interesting study. 

\section{Conclusion}
\label{Conclusion}

The most metal-rich stars in the universe, $0.014\lesssim Z\lesssim 0.10$, are consistently missing from the published stellar yields of AGB stars. The motivation for our research is to fill this missing gap by determining how the chemical contributions of AGB stars vary as metallicity increases. Based on the evolutionary sequences published in \citet{Karakas21}, our research provides the first set of surface abundances and stellar yields with super solar metallicities up to $Z=0.10$ and the first \textit{s}-process yields for metal-rich stars with $Z\geq0.04$. 

Our models indicate that as metallicity increases, TDU efficiency is significantly reduced. Consequently, much of the nucleosynthesis that occurs in the central regions of metal-rich TP-AGB stars remains locked inside the star and is not mixed to the surface. As a result, our models show destruction of $^{12}$C, $^{16}$O and very little production of $^{25}$Mg, $^{26}$Mg, $^{19}$F, $^{26}$Al and $^{22}$Ne. $^{23}$Na scales with higher metallicity, indicating that is mainly secondary production mixed to the surface prior to the TP-AGB. We conclude that the surface abundances of low mass metal-rich stars are primarily governed by FDU and a lack of TDU. 

Metal-rich stars are much less dense than their lower metallicity counterparts due to their higher opacities. This severely impacts the maximum temperatures they can reach at the base of the convective envelope, which allows for solely \textit{partial} hydrogen burning by HBB (if any burning occurs at all). However, we see meaningful production of $^{14}$N, $^{17}$O, $^{13}$C and destruction of $^{15}$N, $^{16}$O and $^{18}$O that deepens with increasing metallicity and mass, regardless of the temperature at the base of the convective envelope. This indicates that in metal-rich intermediate mass AGB stars, the majority of the proton capture nucleosynthesis in their yields is likely a combination of both FDU and SDU with only some contribution from HBB. Accordingly, we have a mix of both primary and secondary isotope production in metal-rich stars, weighted towards secondary production as metallicity increases. 
The chemical contributions of metal-rich stellar populations denote similar trends but are weighted towards the lower mass end of the spectrum. Thus, they are primarily governed by partial hydrogen burning mixed to the surface by FDU, given the majority of stars are those below the mass threshold for SDU. 

Our analysis of the \textit{s}-process in high metallicity intermediate mass stars (for the range $Z=0.04-0.06$) shows there is some production that centres around the first \textit{s}-process process peak at strontium. The magnitude of the peak is small in comparison to lower metallicity models and decreases with increasing metallicity. This is predominantly due to inefficient TDU, the dredged mass of which is then diluted into massive envelopes. We conclude that metal-rich single stars are unlikely candidates for major contributions to the heavy element enrichment of the ISM, but while small-their contributions are non-negligible. 

Stellar yields are one of the theorists’ best methods of refining our understanding of nuclear and mixing processes in the central regions of stars. In this work, we attempt to extend this understanding to the most metal-rich regions in the universe. An obvious extension to our research is to propagate these stellar yields to the galactic scale, by inputting metal-rich yields into galactic chemical evolution (GCE) simulations. GCE simulations allow us to derive a chronology of stellar events that lead us to the current abundances we see in a galaxy \citep{Romano10, Kobayashi11, Kobayashi20}. Work such as \citet{Cote17} have found that regardless of the models’ intricacy, the input stellar yields define the accuracy of the galactic model. 

\section*{Acknowledgements}

We thank the Referee, M. Busso, for his detailed comments and suggestions which have helped to improve the manuscript. We are also grateful to C. Doherty, M. Hampel, J. Lattanzio and R. Willcox for useful discussions. This research was supported by the Australian Research Council Centre of Excellence for All Sky Astrophysics in 3 Dimensions (ASTRO 3D), through project number CE170100013. 

\section*{Data Availability}

The data underlying this article are available in the article and in its online supplementary material.



\bibliographystyle{mnras}
\bibliography{example} 

\begin{thebibliography}{}
\makeatletter
\relax
\def\mn@urlcharsother{\let\do\@makeother \do\$\do\&\do\#\do\^\do\_\do\%\do\~}
\def\mn@doi{\begingroup\mn@urlcharsother \@ifnextchar [ {\mn@doi@}
  {\mn@doi@[]}}
\def\mn@doi@[#1]#2{\def\@tempa{#1}\ifx\@tempa\@empty \href
  {http://dx.doi.org/#2} {doi:#2}\else \href {http://dx.doi.org/#2} {#1}\fi
  \endgroup}
\def\mn@eprint#1#2{\mn@eprint@#1:#2::\@nil}
\def\mn@eprint@arXiv#1{\href {http://arxiv.org/abs/#1} {{\tt arXiv:#1}}}
\def\mn@eprint@dblp#1{\href {http://dblp.uni-trier.de/rec/bibtex/#1.xml}
  {dblp:#1}}
\def\mn@eprint@#1:#2:#3:#4\@nil{\def\@tempa {#1}\def\@tempb {#2}\def\@tempc
  {#3}\ifx \@tempc \@empty \let \@tempc \@tempb \let \@tempb \@tempa \fi \ifx
  \@tempb \@empty \def\@tempb {arXiv}\fi \@ifundefined
  {mn@eprint@\@tempb}{\@tempb:\@tempc}{\expandafter \expandafter \csname
  mn@eprint@\@tempb\endcsname \expandafter{\@tempc}}}

\bibitem[\protect\citeauthoryear{Abia, Hedrosa, Dom{\'\i}nguez  \&
  Straniero}{Abia et~al.}{2017}]{Abia17mixing}
Abia C.,  Hedrosa R.,  Dom{\'\i}nguez I.,   Straniero O.,  2017, Astronomy \&
  Astrophysics, 599, A39

\bibitem[\protect\citeauthoryear{Abia, Cristallo, Cunha, De~Laverny  \&
  Smith}{Abia et~al.}{2019}]{Abia19}
Abia C.,  Cristallo S.,  Cunha K.,  De~Laverny P.,   Smith V.,  2019, Astronomy
  \& Astrophysics, 625, A40

\bibitem[\protect\citeauthoryear{Anders \& Zinner}{Anders \&
  Zinner}{1993}]{Anders93SiC}
Anders E.,  Zinner E.,  1993, Meteoritics, 28, 490

\bibitem[\protect\citeauthoryear{Angelou, Church, Stancliffe, Lattanzio  \&
  Smith}{Angelou et~al.}{2011}]{Angelou11thermohaline}
Angelou G.~C.,  Church R.~P.,  Stancliffe R.~J.,  Lattanzio J.~C.,   Smith
  G.~H.,  2011, The Astrophysical Journal, 728, 79

\bibitem[\protect\citeauthoryear{Angelou, D'Orazi, Constantino, Church,
  Stancliffe  \& Lattanzio}{Angelou et~al.}{2015}]{Angelou15}
Angelou G.~C.,  D'Orazi V.,  Constantino T.~N.,  Church R.~P.,  Stancliffe
  R.~J.,   Lattanzio J.~C.,  2015, Monthly Notices of the Royal Astronomical
  Society, 450, 2423

\bibitem[\protect\citeauthoryear{Arnould}{Arnould}{1987}]{Arnould87}
Arnould M.,  1987, Phil. Trans. R. Soc. Lond., 323, 251

\bibitem[\protect\citeauthoryear{Asplund, Grevesse, Sauval  \& Scott}{Asplund
  et~al.}{2009}]{Asplund09}
Asplund M.,  Grevesse N.,  Sauval A.~J.,   Scott P.,  2009, Ann. Rev. of A \&
  A, 47, 481

\bibitem[\protect\citeauthoryear{Ballero, Kroupa  \& Matteucci}{Ballero
  et~al.}{2007}]{Ballero07}
Ballero S.~K.,  Kroupa P.,   Matteucci F.,  2007, Astronomy \& Astrophysics,
  467, 117

\bibitem[\protect\citeauthoryear{Battino et~al.,}{Battino
  et~al.}{2019}]{Battino19}
Battino U.,  et~al., 2019, Monthly Notices of the Royal Astronomical Society,
  489, 1082

\bibitem[\protect\citeauthoryear{Becker \& Iben}{Becker \&
  Iben}{1979}]{Becker79}
Becker S.,  Iben I.,  1979, ApJ, 232, 831

\bibitem[\protect\citeauthoryear{Bertelli, Girardi, Marigo  \& Nasi}{Bertelli
  et~al.}{2008}]{Bertelli08}
Bertelli G.,  Girardi L.,  Marigo P.,   Nasi E.,  2008, A \& A, 484, 815

\bibitem[\protect\citeauthoryear{Bl\"{o}cker}{Bl\"{o}cker}{1995}]{Blocker95a}
Bl\"{o}cker T.,  1995, A\&AS, 297, 727

\bibitem[\protect\citeauthoryear{Boothroyd \& Sackmann}{Boothroyd \&
  Sackmann}{1988}]{Boothroyd88}
Boothroyd A.~I.,  Sackmann I.,  1988, The Astrophysical Journal, 328, 671

\bibitem[\protect\citeauthoryear{Boyer et~al.,}{Boyer et~al.}{2019}]{Boyer19}
Boyer M.,  et~al., 2019, ApJ, 879, 109

\bibitem[\protect\citeauthoryear{Buntain, Doherty, Lugaro, Lattanzio,
  Stancliffe  \& Karakas}{Buntain et~al.}{2017}]{Buntain17partialmixing}
Buntain J.,  Doherty C.,  Lugaro M.,  Lattanzio J.,  Stancliffe R.,   Karakas
  A.~I.,  2017, Monthly Notices of the Royal Astronomical Society, 471, 824

\bibitem[\protect\citeauthoryear{Busso, Gallino  \& Wasserburg}{Busso
  et~al.}{1999}]{Busso99}
Busso M.,  Gallino R.,   Wasserburg G.,  1999, Annual Review of Astronomy and
  Astrophysics, 37, 239

\bibitem[\protect\citeauthoryear{Busso, Straniero, Gallino  \& Abia}{Busso
  et~al.}{2004}]{Busso04}
Busso M.,  Straniero O.,  Gallino R.,   Abia C.,  2004, Origin and Evolution of
  the Elements, p.~67

\bibitem[\protect\citeauthoryear{Busso, Palmerini, Maiorca, Cristallo,
  Straniero, Abia, Gallino  \& La~Cognata}{Busso
  et~al.}{2010}]{Busso10deepmixing}
Busso M.,  Palmerini S.,  Maiorca E.,  Cristallo S.,  Straniero O.,  Abia C.,
  Gallino R.,   La~Cognata M.,  2010, The Astrophysical Journal Letters, 717,
  L47

\bibitem[\protect\citeauthoryear{Busso, Vescovi, Palmerini, Cristallo  \&
  Antonuccio-Delogu}{Busso et~al.}{2021}]{Busso21}
Busso M.,  Vescovi D.,  Palmerini S.,  Cristallo S.,   Antonuccio-Delogu V.,
  2021, The Astrophysical Journal, 908, 55

\bibitem[\protect\citeauthoryear{Cannon}{Cannon}{1993}]{Cannon93}
Cannon R.~C.,  1993, MNRAS, 263, 817

\bibitem[\protect\citeauthoryear{Carrera, Gallart, Hardy, Aparicio  \&
  Zinn}{Carrera et~al.}{2008}]{Carrera08}
Carrera R.,  Gallart C.,  Hardy E.,  Aparicio A.,   Zinn R.,  2008, The
  Astronomical Journal, 135, 836

\bibitem[\protect\citeauthoryear{Charbonnel \& Lagarde}{Charbonnel \&
  Lagarde}{2010}]{Charbonnel10thermohaline1}
Charbonnel C.,  Lagarde N.,  2010, Astronomy \& Astrophysics, 522, A10

\bibitem[\protect\citeauthoryear{Charbonnel \& Zahn}{Charbonnel \&
  Zahn}{2007}]{Charbonnel07}
Charbonnel C.,  Zahn J.-P.,  2007, Astronomy \& Astrophysics, 467, L15

\bibitem[\protect\citeauthoryear{Charbonnel \& do Nascimento~Jr}{Charbonnel \&
  do~Nascimento~Jr}{1998}]{Charbonnel98}
Charbonnel C.,  do Nascimento~Jr J.~D.,  1998, arXiv preprint astro-ph/9805235

\bibitem[\protect\citeauthoryear{Choi, Dotter, Conroy, Cantiello, Paxton  \&
  Johnson}{Choi et~al.}{2016}]{Choi16}
Choi J.,  Dotter A.,  Conroy C.,  Cantiello M.,  Paxton B.,   Johnson B.~D.,
  2016, ApJ, 823, 102

\bibitem[\protect\citeauthoryear{Cohen, Blakeslee  \& C{\^o}t{\'e}}{Cohen
  et~al.}{2003}]{Cohen03}
Cohen J.~G.,  Blakeslee J.,   C{\^o}t{\'e} P.,  2003, ApJ, 592, 866

\bibitem[\protect\citeauthoryear{C{\^o}t{\'e}, O’Shea, Ritter, Herwig  \&
  Venn}{C{\^o}t{\'e} et~al.}{2017}]{Cote17}
C{\^o}t{\'e} B.,  O’Shea B.~W.,  Ritter C.,  Herwig F.,   Venn K.~A.,  2017,
  ApJ, 835, 128

\bibitem[\protect\citeauthoryear{Cristallo, Gallino, Straniero, Piersanti  \&
  Dominguez}{Cristallo et~al.}{2006}]{Cristallo06}
Cristallo S.,  Gallino R.,  Straniero O.,  Piersanti L.,   Dominguez I.,  2006,
  Mem. Soc. Astron. Ital., 77, 774

\bibitem[\protect\citeauthoryear{Cristallo, Straniero, Gallino, Piersanti,
  Dom{\'\i}nguez  \& Lederer}{Cristallo et~al.}{2009}]{Cristallo09}
Cristallo S.,  Straniero O.,  Gallino R.,  Piersanti L.,  Dom{\'\i}nguez I.,
  Lederer M.,  2009, ApJ, 696, 797

\bibitem[\protect\citeauthoryear{Cristallo, Straniero, Piersanti  \&
  Gobrecht}{Cristallo et~al.}{2015}]{Cristallo15}
Cristallo S.,  Straniero O.,  Piersanti L.,   Gobrecht D.,  2015, ApJ Supp.
  Series, 219, 40

\bibitem[\protect\citeauthoryear{Cyburt et~al.,}{Cyburt
  et~al.}{2010}]{Cyburt10}
Cyburt R.~H.,  et~al., 2010, ApJ Supp. Ser., 189, 240

\bibitem[\protect\citeauthoryear{Davis, Jones  \& Herwig}{Davis
  et~al.}{2019}]{Davis19}
Davis A.,  Jones S.,   Herwig F.,  2019, Monthly Notices of the Royal
  Astronomical Society, 484, 3921

\bibitem[\protect\citeauthoryear{Denissenkov \& Herwig}{Denissenkov \&
  Herwig}{2003}]{Denissenkov03}
Denissenkov P.~A.,  Herwig F.,  2003, ApJ Lett., 590, L99

\bibitem[\protect\citeauthoryear{Denissenkov \& Tout}{Denissenkov \&
  Tout}{2003}]{Denissenkov03grav}
Denissenkov P.~A.,  Tout C.~A.,  2003, Monthly Notices of the Royal
  Astronomical Society, 340, 722

\bibitem[\protect\citeauthoryear{Do, Kerzendorf, Winsor, St{\o}stad, Morris, Lu
   \& Ghez}{Do et~al.}{2015}]{Do15}
Do T.,  Kerzendorf W.,  Winsor N.,  St{\o}stad M.,  Morris M.~R.,  Lu J.~R.,
  Ghez A.~M.,  2015, ApJ, 809, 143

\bibitem[\protect\citeauthoryear{Feltzing \& Chiba}{Feltzing \&
  Chiba}{2013}]{Feltzing13}
Feltzing S.,  Chiba M.,  2013, New Astron. Rev., 57, 80

\bibitem[\protect\citeauthoryear{Fishlock, Karakas, Lugaro  \& Yong}{Fishlock
  et~al.}{2014}]{Fishlock14}
Fishlock C.~K.,  Karakas A.~I.,  Lugaro M.,   Yong D.,  2014, The Astrophysical
  Journal, 797, 44

\bibitem[\protect\citeauthoryear{Forestini \& Charbonnel}{Forestini \&
  Charbonnel}{1997}]{Forestini97}
Forestini M.,  Charbonnel C.,  1997, A\&AS Supp. Series, 123, 241

\bibitem[\protect\citeauthoryear{Forestini, Arnould  \& Paulus}{Forestini
  et~al.}{1991}]{Forestini91}
Forestini M.,  Arnould M.,   Paulus G.,  1991, A\&A, 252, 597

\bibitem[\protect\citeauthoryear{Frantsman}{Frantsman}{1989}]{Frantsman89}
Frantsman Y.~L.,  1989, Sov. Astr., 33, 565

\bibitem[\protect\citeauthoryear{Frost \& Lattanzio}{Frost \&
  Lattanzio}{1996}]{Frost96}
Frost C.,  Lattanzio J.,  1996, ApJ, 473, 383

\bibitem[\protect\citeauthoryear{Gallino, Busso, Picchio  \& Raiteri}{Gallino
  et~al.}{1990}]{Gallino90SiC}
Gallino R.,  Busso M.,  Picchio G.,   Raiteri C.,  1990, Nature, 348, 298

\bibitem[\protect\citeauthoryear{Gallino, Arlandini, Busso, Lugaro, Travaglio,
  Straniero, Chieffi  \& Limongi}{Gallino et~al.}{1998}]{Gallino98}
Gallino R.,  Arlandini C.,  Busso M.,  Lugaro M.,  Travaglio C.,  Straniero O.,
   Chieffi A.,   Limongi M.,  1998, The Astrophysical Journal, 497, 388

\bibitem[\protect\citeauthoryear{Gibson, Fenner, Renda, Kawata  \& Lee}{Gibson
  et~al.}{2003}]{Gibson03}
Gibson B.~K.,  Fenner Y.,  Renda A.,  Kawata D.,   Lee H.-c.,  2003,
  Publications of the Astronomical Society of Australia, 20, 401

\bibitem[\protect\citeauthoryear{Goriely \& Siess}{Goriely \&
  Siess}{2004}]{Goriely04}
Goriely S.,  Siess L.,  2004, Astronomy \& Astrophysics, 421, L25

\bibitem[\protect\citeauthoryear{Hale, Champagne, Iliadis, Hansper, Powell  \&
  Blackmon}{Hale et~al.}{2001}]{Hale01}
Hale S.,  Champagne A.,  Iliadis C.,  Hansper V.,  Powell D.,   Blackmon J.,
  2001, Phys. Rev. C, 65, 015801

\bibitem[\protect\citeauthoryear{Heck, Marhas, Hoppe, Gallino, Baur  \&
  Wieler}{Heck et~al.}{2007}]{Heck07SiC}
Heck P.~R.,  Marhas K.~K.,  Hoppe P.,  Gallino R.,  Baur H.,   Wieler R.,
  2007, The Astrophysical Journal, 656, 1208

\bibitem[\protect\citeauthoryear{Henkel, Karakas  \& Lattanzio}{Henkel
  et~al.}{2017}]{Henkel2017}
Henkel K.,  Karakas A.~I.,   Lattanzio J.~C.,  2017, Monthly Notices of the
  Royal Astronomical Society, 469, 4600

\bibitem[\protect\citeauthoryear{Herwig}{Herwig}{2000}]{Herwig00}
Herwig F.,  2000, arXiv preprint astro-ph/0007139

\bibitem[\protect\citeauthoryear{Herwig}{Herwig}{2004a}]{Herwig04}
Herwig F.,  2004a, ApJ Supp. Series, 155, 651

\bibitem[\protect\citeauthoryear{Herwig}{Herwig}{2004b}]{Herwig2004dredge}
Herwig F.,  2004b, The Astrophysical Journal, 605, 425

\bibitem[\protect\citeauthoryear{Herwig}{Herwig}{2005}]{Herwig05}
Herwig F.,  2005, Annu. Rev. Astron. Astrophys., 43, 435

\bibitem[\protect\citeauthoryear{Herwig, Austin  \& Lattanzio}{Herwig
  et~al.}{2006}]{Herwig06nuclear}
Herwig F.,  Austin S.~M.,   Lattanzio J.~C.,  2006, Physical Review C, 73,
  025802

\bibitem[\protect\citeauthoryear{Iben~Jr}{Iben~Jr}{1967a}]{Iben67}
Iben~Jr I.,  1967a, Annual Review of Astronomy and Astrophysics, 5, 571

\bibitem[\protect\citeauthoryear{Iben~Jr}{Iben~Jr}{1967b}]{Iben67stellar}
Iben~Jr I.,  1967b, The Astrophysical Journal, 147, 624

\bibitem[\protect\citeauthoryear{Izzard, Dray, Karakas, Lugaro  \& Tout}{Izzard
  et~al.}{2006}]{Izzard06}
Izzard R.~G.,  Dray L.~M.,  Karakas A.~I.,  Lugaro M.,   Tout C.~A.,  2006,
  Astronomy \& Astrophysics, 460, 565

\bibitem[\protect\citeauthoryear{Izzard, Lugaro, Karakas, Iliadis  \& van
  Raai}{Izzard et~al.}{2007}]{Izzard07}
Izzard R.~G.,  Lugaro M.,  Karakas A.~I.,  Iliadis C.,   van Raai M.,  2007,
  A\&A, 466, 641

\bibitem[\protect\citeauthoryear{Jorissen, Smith  \& Lambert}{Jorissen
  et~al.}{1992}]{Jorissen92}
Jorissen A.,  Smith V.,   Lambert D.,  1992, A\&A, 261, 164

\bibitem[\protect\citeauthoryear{Kamath, Karakas  \& Wood}{Kamath
  et~al.}{2012}]{Kamath12}
Kamath D.,  Karakas A.~I.,   Wood P.~R.,  2012, The Astrophysical Journal, 746,
  20

\bibitem[\protect\citeauthoryear{K{\"a}ppeler, Gallino, Bisterzo  \&
  Aoki}{K{\"a}ppeler et~al.}{2011}]{Kappeler11}
K{\"a}ppeler F.,  Gallino R.,  Bisterzo S.,   Aoki W.,  2011, Rev. Mod. Phys.,
  83, 157

\bibitem[\protect\citeauthoryear{Karakas}{Karakas}{2010}]{Karakas10Up}
Karakas A.~I.,  2010, MNRAS, 403, 1413

\bibitem[\protect\citeauthoryear{Karakas}{Karakas}{2014}]{Karakas14He}
Karakas A.~I.,  2014, MNRAS, 445, 347

\bibitem[\protect\citeauthoryear{Karakas \& Lattanzio}{Karakas \&
  Lattanzio}{2003a}]{Karakas03}
Karakas A.~I.,  Lattanzio J.~C.,  2003a, Publ. Astron. Soc. Aus., 20, 279

\bibitem[\protect\citeauthoryear{Karakas \& Lattanzio}{Karakas \&
  Lattanzio}{2003b}]{Karakas03neon}
Karakas A.~I.,  Lattanzio J.~C.,  2003b, Publications of the Astronomical
  Society of Australia, 20, 393

\bibitem[\protect\citeauthoryear{Karakas \& Lattanzio}{Karakas \&
  Lattanzio}{2007}]{Karakas07}
Karakas A.,  Lattanzio J.~C.,  2007, Pub. Astron. Soc. Aus., 24, 103

\bibitem[\protect\citeauthoryear{Karakas \& Lattanzio}{Karakas \&
  Lattanzio}{2014}]{Karakas14}
Karakas A.~I.,  Lattanzio J.~C.,  2014, Pub. Astron. Soc. Aus., 31

\bibitem[\protect\citeauthoryear{Karakas \& Lugaro}{Karakas \&
  Lugaro}{2016}]{Karakas16}
Karakas A.~I.,  Lugaro M.,  2016, ApJ, 825, 26

\bibitem[\protect\citeauthoryear{Karakas, Lattanzio  \& Pols}{Karakas
  et~al.}{2002}]{Karakas02}
Karakas A.~I.,  Lattanzio J.,   Pols O.~R.,  2002, Publ. Astron. Soc. Aus., 19,
  515

\bibitem[\protect\citeauthoryear{Karakas, Lugaro, Wiescher, G{\"o}rres  \&
  Ugalde}{Karakas et~al.}{2006}]{Karakas06}
Karakas A.~I.,  Lugaro M.~A.,  Wiescher M.,  G{\"o}rres J.,   Ugalde C.,  2006,
  ApJ, 643, 471

\bibitem[\protect\citeauthoryear{Karakas, van Raai, Lugaro, Sterling  \&
  Dinerstein}{Karakas et~al.}{2009}]{Karakas09}
Karakas A.~I.,  van Raai M.~A.,  Lugaro M.,  Sterling N.~C.,   Dinerstein
  H.~L.,  2009, The Astrophysical Journal, 690, 1130

\bibitem[\protect\citeauthoryear{Karakas, Campbell  \& Stancliffe}{Karakas
  et~al.}{2010}]{Karakas10mixing}
Karakas A.~I.,  Campbell S.~W.,   Stancliffe R.~J.,  2010, The Astrophysical
  Journal, 713, 374

\bibitem[\protect\citeauthoryear{Karakas, Lugaro, Carlos, Cseh, Kamath  \&
  Garc{\'\i}a-Hern{\'a}ndez}{Karakas et~al.}{2018}]{Karakas18}
Karakas A.~I.,  Lugaro M.,  Carlos M.,  Cseh B.,  Kamath D.,
  Garc{\'\i}a-Hern{\'a}ndez D.,  2018, MNRAS, 477, 421

\bibitem[\protect\citeauthoryear{Karakas, Cinquegrana  \& Joyce}{Karakas
  et~al.}{2021}]{Karakas21}
Karakas A.,  Cinquegrana G.,   Joyce M.,  2021, The most metal-rich asymptotic
  giant branch stars (\mn@eprint {arXiv} {2111.01308})

\bibitem[\protect\citeauthoryear{Kobayashi, Umeda, Nomoto, Tominaga  \&
  Ohkubo}{Kobayashi et~al.}{2006}]{Kobayashi06}
Kobayashi C.,  Umeda H.,  Nomoto K.,  Tominaga N.,   Ohkubo T.,  2006, ApJ,
  653, 1145

\bibitem[\protect\citeauthoryear{Kobayashi, Karakas  \& Umeda}{Kobayashi
  et~al.}{2011}]{Kobayashi11}
Kobayashi C.,  Karakas A.~I.,   Umeda H.,  2011, Monthly Notices of the Royal
  Astronomical Society, 414, 3231

\bibitem[\protect\citeauthoryear{Kobayashi, Karakas  \& Lugaro}{Kobayashi
  et~al.}{2020}]{Kobayashi20}
Kobayashi C.,  Karakas A.~I.,   Lugaro M.,  2020, The Astrophysical Journal,
  900, 179

\bibitem[\protect\citeauthoryear{Kroupa}{Kroupa}{2001}]{Kroupa01}
Kroupa P.,  2001, Monthly Notices of the Royal Astronomical Society, 322, 231

\bibitem[\protect\citeauthoryear{Kroupa, Tout  \& Gilmore}{Kroupa
  et~al.}{1993}]{Kroupa93}
Kroupa P.,  Tout C.~A.,   Gilmore G.,  1993, Monthly Notices of the Royal
  Astronomical Society, 262, 545

\bibitem[\protect\citeauthoryear{La~Cognata et~al.,}{La~Cognata
  et~al.}{2011}]{LaCognata11}
La~Cognata M.,  et~al., 2011, ApJ Lett., 739, L54

\bibitem[\protect\citeauthoryear{Lagarde, Charbonnel, Decressin  \&
  Hagelberg}{Lagarde et~al.}{2011}]{Lagarde11thermohaline2}
Lagarde N.,  Charbonnel C.,  Decressin T.,   Hagelberg J.,  2011, Astronomy \&
  Astrophysics, 536, A28

\bibitem[\protect\citeauthoryear{Lagarde, Romano, Charbonnel, Tosi, Chiappini
  \& Matteucci}{Lagarde et~al.}{2012}]{Lagarde12}
Lagarde N.,  Romano D.,  Charbonnel C.,  Tosi M.,  Chiappini C.,   Matteucci
  F.,  2012, Astronomy \& Astrophysics, 542, A62

\bibitem[\protect\citeauthoryear{Lattanzio}{Lattanzio}{2003}]{Lattanzio03}
Lattanzio J.,  2003, in Symp.-Int. Astron. Un.. pp 73--81

\bibitem[\protect\citeauthoryear{Lattanzio \& Forestini}{Lattanzio \&
  Forestini}{1999}]{Lattanzio99}
Lattanzio J.,  Forestini M.,  1999, in Symp.Int. Astro. Un.. pp 31--40

\bibitem[\protect\citeauthoryear{Lattanzio, Frost, Cannon  \& Wood}{Lattanzio
  et~al.}{1996}]{Lattanzio96}
Lattanzio J.,  Frost C.,  Cannon R.,   Wood P.,  1996, Memorie della
  Societ{\`a} astronomica italiana, 67, 729

\bibitem[\protect\citeauthoryear{Lattanzio, Siess, Church, Angelou, Stancliffe,
  Doherty, Stephen  \& Campbell}{Lattanzio et~al.}{2015}]{Lattanzio15}
Lattanzio J.~C.,  Siess L.,  Church R.~P.,  Angelou G.,  Stancliffe R.~J.,
  Doherty C.~L.,  Stephen T.,   Campbell S.~W.,  2015, Monthly Notices of the
  Royal Astronomical Society, 446, 2673

\bibitem[\protect\citeauthoryear{L{\'e}pine et~al.,}{L{\'e}pine
  et~al.}{2011}]{Lepine11}
L{\'e}pine J.~R.,  et~al., 2011, MNRAS, 417, 698

\bibitem[\protect\citeauthoryear{Lugaro}{Lugaro}{2016}]{Lugaro16}
Lugaro M.,  2016, in Journal of Physics: Conference Series. p. 012003

\bibitem[\protect\citeauthoryear{Lugaro, Davis, Gallino, Pellin, Straniero  \&
  K{\"a}ppeler}{Lugaro et~al.}{2003}]{Lugaro03SiC}
Lugaro M.,  Davis A.~M.,  Gallino R.,  Pellin M.~J.,  Straniero O.,
  K{\"a}ppeler F.,  2003, The Astrophysical Journal, 593, 486

\bibitem[\protect\citeauthoryear{Lugaro, Ugalde, Karakas, G{\"o}rres, Wiescher,
  Lattanzio  \& Cannon}{Lugaro et~al.}{2004}]{Lugaro04reaction}
Lugaro M.,  Ugalde C.,  Karakas A.~I.,  G{\"o}rres J.,  Wiescher M.,  Lattanzio
  J.~C.,   Cannon R.~C.,  2004, The Astrophysical Journal, 615, 934

\bibitem[\protect\citeauthoryear{Lugaro, Karakas, Stancliffe  \& Rijs}{Lugaro
  et~al.}{2012}]{Lugaro12}
Lugaro M.,  Karakas A.~I.,  Stancliffe R.~J.,   Rijs C.,  2012, ApJ, 747, 2

\bibitem[\protect\citeauthoryear{Lugaro, Karakas, Pignatari  \& Doherty}{Lugaro
  et~al.}{2016}]{Lugaro16stellar}
Lugaro M.,  Karakas A.~I.,  Pignatari M.,   Doherty C.~L.,  2016, Proceedings
  of the International Astronomical Union, 12, 86

\bibitem[\protect\citeauthoryear{Lugaro, Karakas, Pet{\H{o}}  \& Plachy}{Lugaro
  et~al.}{2018}]{Lugaro18meteoritic}
Lugaro M.,  Karakas A.~I.,  Pet{\H{o}} M.,   Plachy E.,  2018, Geochimica et
  Cosmochimica Acta, 221, 6

\bibitem[\protect\citeauthoryear{Lugaro et~al.,}{Lugaro
  et~al.}{2020}]{Lugaro20origin}
Lugaro M.,  et~al., 2020, The Astrophysical Journal, 898, 96

\bibitem[\protect\citeauthoryear{Maoz, Mannucci  \& Nelemans}{Maoz
  et~al.}{2014}]{Maoz14}
Maoz D.,  Mannucci F.,   Nelemans G.,  2014, Annual Review of Astronomy and
  Astrophysics, 52, 107

\bibitem[\protect\citeauthoryear{Marigo}{Marigo}{2001}]{Marigo01}
Marigo P.,  2001, A\&AS, 370, 194

\bibitem[\protect\citeauthoryear{Marigo, Girardi, Garching, Padova  \&
  Observ.}{Marigo et~al.}{1999}]{Marigo99}
Marigo P.,  Girardi L.,  Garching A. W.~M.,  Padova D.~A.,   Observ. P.,  1999,
  Astronomy and Astrophysics, 344, 123

\bibitem[\protect\citeauthoryear{Marigo et~al.,}{Marigo
  et~al.}{2017}]{Marigo17}
Marigo P.,  et~al., 2017, ApJ, 835, 77

\bibitem[\protect\citeauthoryear{Marks, Kroupa, Dabringhausen  \&
  Pawlowski}{Marks et~al.}{2012}]{Marks12}
Marks M.,  Kroupa P.,  Dabringhausen J.,   Pawlowski M.~S.,  2012, Monthly
  Notices of the Royal Astronomical Society, 422, 2246

\bibitem[\protect\citeauthoryear{Meynet, Mowlavi  \& Maeder}{Meynet
  et~al.}{2006}]{Meynet06}
Meynet G.,  Mowlavi N.,   Maeder A.,  2006, ArX. Pre. astro-ph/0611261

\bibitem[\protect\citeauthoryear{Mowlavi}{Mowlavi}{1999}]{Mowlavi99sodium}
Mowlavi N.,  1999, arXiv preprint astro-ph/9910542

\bibitem[\protect\citeauthoryear{Mowlavi, Meynet, Maeder, Schaerer  \&
  Charbonnel}{Mowlavi et~al.}{1998}]{Mowlavi98}
Mowlavi N.,  Meynet G.,  Maeder A.,  Schaerer D.,   Charbonnel C.,  1998, ArX.
  Pre. astro-ph/9804155

\bibitem[\protect\citeauthoryear{Nanni, Bressan, Marigo  \& Girardi}{Nanni
  et~al.}{2014}]{Nanni14evolution}
Nanni A.,  Bressan A.,  Marigo P.,   Girardi L.,  2014, Monthly Notices of the
  Royal Astronomical Society, 438, 2328

\bibitem[\protect\citeauthoryear{Nittler \& Ciesla}{Nittler \&
  Ciesla}{2016}]{Nittler16grains}
Nittler L.~R.,  Ciesla F.,  2016, Annual Review of Astronomy and Astrophysics,
  54, 53

\bibitem[\protect\citeauthoryear{Nomoto, Kobayashi  \& Tominaga}{Nomoto
  et~al.}{2013}]{Nomoto13}
Nomoto K.,  Kobayashi C.,   Tominaga N.,  2013, Annual Review of Astronomy and
  Astrophysics, 51, 457

\bibitem[\protect\citeauthoryear{Norgaard}{Norgaard}{1980}]{Norgaard80}
Norgaard H.,  1980, The Astrophysical Journal, 236, 895

\bibitem[\protect\citeauthoryear{Origlia, Valenti, Rich  \& Ferraro}{Origlia
  et~al.}{2006}]{Origlia06}
Origlia L.,  Valenti E.,  Rich R.~M.,   Ferraro F.~R.,  2006, ApJ, 646, 499

\bibitem[\protect\citeauthoryear{Palmerini, Cristallo, Piersanti, Vescovi  \&
  Busso}{Palmerini et~al.}{2021}]{Palmerini21grains}
Palmerini S.,  Cristallo S.,  Piersanti L.,  Vescovi D.,   Busso M.,  2021,
  Universe, 7, 175

\bibitem[\protect\citeauthoryear{Perea-Calder{\'o}n, Garc{\'\i}a-Hern{\'a}ndez,
  Garc{\'\i}a-Lario, Szczerba  \& Bobrowsky}{Perea-Calder{\'o}n
  et~al.}{2009}]{Perea09}
Perea-Calder{\'o}n J.,  Garc{\'\i}a-Hern{\'a}ndez D.,  Garc{\'\i}a-Lario P.,
  Szczerba R.,   Bobrowsky M.,  2009, Astronomy \& Astrophysics, 495, L5

\bibitem[\protect\citeauthoryear{Pols \& Tout}{Pols \&
  Tout}{2001}]{Pols01thermal}
Pols O.~R.,  Tout C.~A.,  2001, Memorie della Societa Astronomica Italiana, 72,
  299

\bibitem[\protect\citeauthoryear{Prantzos, Abia, Limongi, Chieffi  \&
  Cristallo}{Prantzos et~al.}{2018}]{Prantzos18}
Prantzos N.,  Abia C.,  Limongi M.,  Chieffi A.,   Cristallo S.,  2018, Monthly
  Notices of the Royal Astronomical Society, 476, 3432

\bibitem[\protect\citeauthoryear{Reimers}{Reimers}{1975}]{Reimers75}
Reimers D.,  1975, in , Problems in stellar atmospheres and envelopes.
Springer, pp 229--256

\bibitem[\protect\citeauthoryear{Romano, Karakas, Tosi  \& Matteucci}{Romano
  et~al.}{2010}]{Romano10}
Romano D.,  Karakas A.~I.,  Tosi M.,   Matteucci F.,  2010, A \& A, 522, A32

\bibitem[\protect\citeauthoryear{Rood, Steigman  \& Tinsley}{Rood
  et~al.}{1976}]{Rood76}
Rood R.~T.,  Steigman G.,   Tinsley B.~M.,  1976, ApJ, 207, L57

\bibitem[\protect\citeauthoryear{Rood, Bania  \& Wilson}{Rood
  et~al.}{1984}]{Rood84}
Rood R.~T.,  Bania T.,   Wilson T.,  1984, The Astrophysical Journal, 280, 629

\bibitem[\protect\citeauthoryear{Salpeter}{Salpeter}{1955}]{Salpeter55}
Salpeter E.~E.,  1955, ApJ, 121, 161

\bibitem[\protect\citeauthoryear{Schultheis, Lan{\c{c}}on, Omont, Schuller  \&
  Ojha}{Schultheis et~al.}{2003}]{Schultheis03}
Schultheis M.,  Lan{\c{c}}on A.,  Omont A.,  Schuller F.,   Ojha D.,  2003,
  Astronomy \& Astrophysics, 405, 531

\bibitem[\protect\citeauthoryear{Schultheis et~al.,}{Schultheis
  et~al.}{2020}]{Schultheis20}
Schultheis M.,  et~al., 2020, Astronomy \& Astrophysics, 642, A81

\bibitem[\protect\citeauthoryear{Sestito, Randich, Mermilliod  \&
  Pallavicini}{Sestito et~al.}{2003}]{Sestito03}
Sestito P.,  Randich S.,  Mermilliod J.-C.,   Pallavicini R.,  2003, A \& A,
  407, 289

\bibitem[\protect\citeauthoryear{Siess}{Siess}{2010}]{Siess10}
Siess L.,  2010, Memorie della Societa Astronomica Italiana, 81, 980

\bibitem[\protect\citeauthoryear{Slemer et~al.,}{Slemer
  et~al.}{2016}]{Slemer16}
Slemer A.,  et~al., 2016, MNRAS, 465, 4817

\bibitem[\protect\citeauthoryear{Smith \& Lambert}{Smith \&
  Lambert}{1989}]{Smith89Li}
Smith V.~V.,  Lambert D.~L.,  1989, The Astrophysical Journal, 345, L75

\bibitem[\protect\citeauthoryear{Smith \& Lambert}{Smith \&
  Lambert}{1990}]{Smith90Li}
Smith V.~V.,  Lambert D.~L.,  1990, The Astrophysical Journal, 361, L69

\bibitem[\protect\citeauthoryear{Sneden, Cowan  \& Gallino}{Sneden
  et~al.}{2008}]{Sneden08}
Sneden C.,  Cowan J.~J.,   Gallino R.,  2008, Annu. Rev. Astron. Astrophys.,
  46, 241

\bibitem[\protect\citeauthoryear{Stancliffe}{Stancliffe}{2010}]{Stancliffe10}
Stancliffe R.~J.,  2010, Monthly Notices of the Royal Astronomical Society,
  403, 505

\bibitem[\protect\citeauthoryear{Straniero, Gallino, Busso, Chiefei, Raiteri,
  Limongi  \& Salaris}{Straniero et~al.}{1995}]{Straniero95}
Straniero O.,  Gallino R.,  Busso M.,  Chiefei A.,  Raiteri C.,  Limongi M.,
  Salaris M.,  1995, The Astrophysical Journal, 440, L85

\bibitem[\protect\citeauthoryear{Straniero, Dominguez, Cristallo  \&
  Gallino}{Straniero et~al.}{2003}]{Straniero03}
Straniero O.,  Dominguez I.,  Cristallo S.,   Gallino R.,  2003, Pub. Astron.
  Soc. Aus., 20, 389

\bibitem[\protect\citeauthoryear{Tinsley}{Tinsley}{1980}]{Tinsley80}
Tinsley B.~M.,  1980, Fund. of cosm. phys., 5, 287

\bibitem[\protect\citeauthoryear{Travaglio, Randich, Galli, Lattanzio, Elliott,
  Forestini  \& Ferrini}{Travaglio et~al.}{2001}]{Travaglio01Lith}
Travaglio C.,  Randich S.,  Galli D.,  Lattanzio J.,  Elliott L.~M.,  Forestini
  M.,   Ferrini F.,  2001, The Astrophysical Journal, 559, 909

\bibitem[\protect\citeauthoryear{Travaglio, Gallino, Rauscher, R{\"o}pke  \&
  Hillebrandt}{Travaglio et~al.}{2015}]{Travaglio15}
Travaglio C.,  Gallino R.,  Rauscher T.,  R{\"o}pke F.~K.,   Hillebrandt W.,
  2015, The Astrophysical Journal, 799, 54

\bibitem[\protect\citeauthoryear{Trippella, Busso, Palmerini, Maiorca  \&
  Nucci}{Trippella et~al.}{2016}]{Trippella16}
Trippella O.,  Busso M.,  Palmerini S.,  Maiorca E.,   Nucci M.,  2016, The
  Astrophysical Journal, 818, 125

\bibitem[\protect\citeauthoryear{Twarog, Anthony-Twarog  \& De~Lee}{Twarog
  et~al.}{2003}]{Twarog03}
Twarog B.~A.,  Anthony-Twarog B.~J.,   De~Lee N.,  2003, ApJ, 125, 1383

\bibitem[\protect\citeauthoryear{Ulrich}{Ulrich}{1972}]{Ulrich1972}
Ulrich R.~K.,  1972, The Astrophysical Journal, 172, 165

\bibitem[\protect\citeauthoryear{Van~Raai, Lugaro, Karakas  \&
  Iliadis}{Van~Raai et~al.}{2008}]{Van08}
Van~Raai M.,  Lugaro M.,  Karakas A.~I.,   Iliadis C.,  2008, Astronomy \&
  Astrophysics, 478, 521

\bibitem[\protect\citeauthoryear{Vassiliadis \& Wood}{Vassiliadis \&
  Wood}{1993}]{Vassiliadis93}
Vassiliadis E.,  Wood P.,  1993, ApJ, 413, 641

\bibitem[\protect\citeauthoryear{Vaughan, Davies, Zieleniewski  \&
  Houghton}{Vaughan et~al.}{2018}]{Vaughan18}
Vaughan S.~P.,  Davies R.~L.,  Zieleniewski S.,   Houghton R.~C.,  2018, MNRAS,
  479, 2443

\bibitem[\protect\citeauthoryear{Ventura \& D'Antona}{Ventura \&
  D'Antona}{2005a}]{Ventura05first}
Ventura P.,  D'Antona F.,  2005a, Astronomy \& Astrophysics, 431, 279

\bibitem[\protect\citeauthoryear{Ventura \& D'Antona}{Ventura \&
  D'Antona}{2005b}]{Ventura05}
Ventura P.,  D'Antona F.,  2005b, A \& A, 439, 1075

\bibitem[\protect\citeauthoryear{Ventura et~al.,}{Ventura
  et~al.}{2012}]{Ventura12dust}
Ventura P.,  et~al., 2012, Monthly Notices of the Royal Astronomical Society,
  424, 2345

\bibitem[\protect\citeauthoryear{Ventura, Di~Criscienzo, Carini  \&
  D’Antona}{Ventura et~al.}{2013}]{Ventura13}
Ventura P.,  Di~Criscienzo M.,  Carini R.,   D’Antona F.,  2013, MNRAS, 431,
  3642

\bibitem[\protect\citeauthoryear{Ventura, Stanghellini, Dell'Agli,
  Garc{\'\i}a-Hern{\'a}ndez  \& Di~Criscienzo}{Ventura
  et~al.}{2015}]{Ventura15}
Ventura P.,  Stanghellini L.,  Dell'Agli F.,  Garc{\'\i}a-Hern{\'a}ndez D.,
  Di~Criscienzo M.,  2015, Monthly Notices of the Royal Astronomical Society,
  452, 3679

\bibitem[\protect\citeauthoryear{Ventura, Karakas, Dell'Agli,
  Garc{\'\i}a-Hern{\'a}ndez  \& Guzman-Ramirez}{Ventura
  et~al.}{2018}]{Ventura18}
Ventura P.,  Karakas A.,  Dell'Agli F.,  Garc{\'\i}a-Hern{\'a}ndez D.,
  Guzman-Ramirez L.,  2018, Monthly Notices of the Royal Astronomical Society,
  475, 2282

\bibitem[\protect\citeauthoryear{Ventura, Dell’Agli, Lugaro, Romano, Tailo
  \& Yag{\"u}e}{Ventura et~al.}{2020}]{Ventura20}
Ventura P.,  Dell’Agli F.,  Lugaro M.,  Romano D.,  Tailo M.,   Yag{\"u}e A.,
   2020, Astronomy \& Astrophysics, 641, A103

\bibitem[\protect\citeauthoryear{Vescovi, Cristallo, Busso  \& Liu}{Vescovi
  et~al.}{2020}]{Vescovi20magnetic}
Vescovi D.,  Cristallo S.,  Busso M.,   Liu N.,  2020, The Astrophysical
  Journal Letters, 897, L25

\bibitem[\protect\citeauthoryear{Vescovi, Cristallo, Palmerini, Abia  \&
  Busso}{Vescovi et~al.}{2021}]{Vescovi21}
Vescovi D.,  Cristallo S.,  Palmerini S.,  Abia C.,   Busso M.,  2021, arXiv
  preprint arXiv:2106.08241

\bibitem[\protect\citeauthoryear{Wallerstein \& Knapp}{Wallerstein \&
  Knapp}{1998}]{Wallerstein98}
Wallerstein G.,  Knapp G.~R.,  1998, Annual Review of Astronomy and
  Astrophysics, 36, 369

\bibitem[\protect\citeauthoryear{Weiss \& Ferguson}{Weiss \&
  Ferguson}{2009}]{Weiss09}
Weiss A.,  Ferguson J.~W.,  2009, A\&AS, 508, 1343

\bibitem[\protect\citeauthoryear{Wood \& Zarro}{Wood \& Zarro}{1981}]{Wood81}
Wood P.,  Zarro D.,  1981, ApJ, 247, 247

\bibitem[\protect\citeauthoryear{Yoon, Langer  \& Van Der~Sluys}{Yoon
  et~al.}{2004}]{Yoon04}
Yoon S.-C.,  Langer N.,   Van Der~Sluys M.,  2004, Astronomy \& Astrophysics,
  425, 207

\bibitem[\protect\citeauthoryear{Zhang, Romano, Ivison, Papadopoulos  \&
  Matteucci}{Zhang et~al.}{2018}]{Zhang18}
Zhang Z.-Y.,  Romano D.,  Ivison R.,  Papadopoulos P.~P.,   Matteucci F.,
  2018, Nature, 558, 260

\bibitem[\protect\citeauthoryear{Zinner}{Zinner}{2014}]{Zinner14grains}
Zinner E.,  2014, Meteorites and cosmochemical processes, 1, 181

\makeatother
\end{thebibliography}




\appendix

\begin{figure}
	\includegraphics[width=\columnwidth]{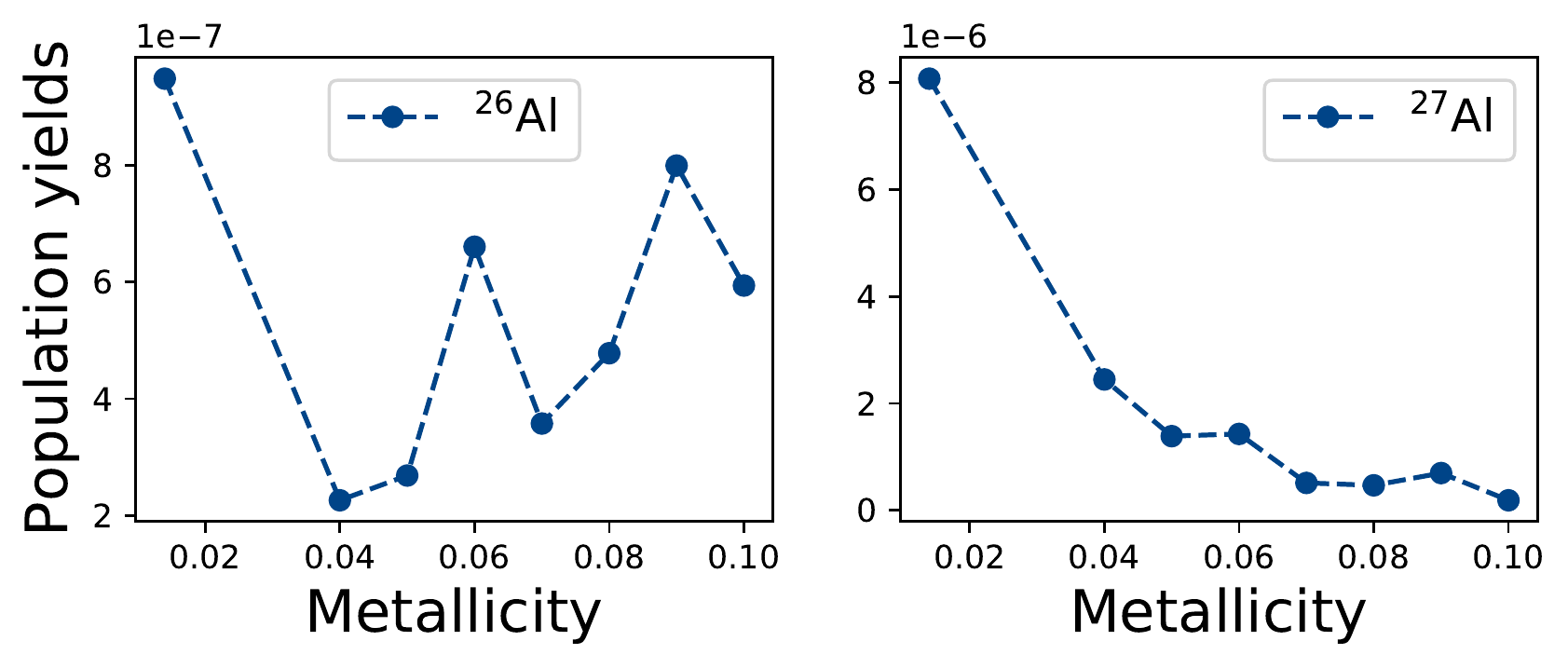}
    \caption{Population yields. Many of the same trends are observed in these plots as we found in sub section \ref{stellaryields}, weighted to a lower mass regime.}
    \label{fig:stellargen2}
\end{figure}

\bsp	
\label{lastpage}
\end{document}